\documentclass[a4paper,11pt]{article}
\usepackage{physics}
\usepackage{jcappub}
\usepackage{pgfplots}
\pgfplotsset{compat=1.18}

\usepackage{amsmath,amssymb, amsthm,amstext,comment,lipsum,mathtools,placeins}
\usepackage{natbib}
\usepackage{graphicx}
\usepackage{xcolor}
\usepackage{array, enumerate}
\usepackage{bm}
\usepackage{multirow}
\usepackage{tikz}
\usetikzlibrary{arrows.meta, decorations.markings}

\newcommand{\rmd}{{\rm d}}
\newcommand{\refEq}[1]{Eq.~(\ref{eq:#1})}
          
\newcommand{\reffig}[1]{Fig.~\ref{fig:#1}}          
\newcommand{\refsec}[1]{Sec.~\ref{sec:#1}}
\newcommand{\refapp}[1]{App.~\ref{app:#1}}
\newcommand{\reftab}[1]{Tab.~\ref{tab:#1}}

\begin{document}

\title{Polarization-shape alignment of IllustrisTNG star-forming galaxies}

\author[a, b]{Rui Zhou,}
\author[b]{Liang Dai,}
\author[c]{Junwu Huang,}
\author[b]{Weichen Winston Yin,}
\author[d,b,e]{and Simone Ferraro}

\affiliation[a]{Department of Physics, Xi'an Jiaotong University, Xi'an, Shaanxi, 710049, China}
\affiliation[b]{Department of Physics, University of California, 366 Physics North MC 7300, Berkeley, CA. 94720, USA}
\affiliation[c]{Perimeter Institute for Theoretical Physics, 31 Caroline St.~N., Waterloo, Ontario N2L 2Y5, Canada}
\affiliation[d]{Physics Division, Lawrence Berkeley National Laboratory, Berkeley, CA 94720, USA}
\affiliation[e]{Berkeley Center for Cosmological Physics, Department of Physics, University of California, Berkeley, CA 94720, USA}

\emailAdd{rzhou@stu.xjtu.edu.cn}
\emailAdd{liangdai@berkeley.edu}
\emailAdd{jhuang@perimeterinstitute.ca}

\abstract{
In star-forming disk galaxies, the radio continuum emission ($1$--$10\,$GHz) powered by star formation has an integrated polarization direction imperfectly aligned with the apparent disk minor axis. This polarization-shape alignment effect was previously observed in a small sample of local spirals. If this is prevalent for disk galaxies out to cosmological redshifts, novel measurements of cosmic birefringence and cosmic shear will be enabled by leveraging radio continuum surveys such as the Square Kilometre Array synergized with galaxy shape measurements. We calculate the polarization-shape misalignment angle for star-forming galaxies in the \texttt{IllustrisTNG50} simulation at $0 < z < 2$, assuming that additional polarized radio emission from an active galactic nucleus is negligible in at least a sizable fraction of the star-forming galaxies. The alignment found for $z=0$ is consistent with local spiral data, but significantly deteriorates as redshift increases. Moreover, it degrades toward lower frequencies due to internal Faraday depolarization. Thanks to cosmic redshifting, observing higher-$z$ galaxies at a fixed frequency greatly mitigates degradation due to reduced Faraday depolarization at the source-frame frequency. We present analytic fits to the non-Gaussian misalignment angle distribution, and evaluate Fisher information per galaxy for measuring a polarization rotation angle induced by cosmic birefringence. For observation at 4.8 GHz, the effective root-mean-square misalignment angle $\sigma_{\alpha,{\rm eff}}$ is $18^\circ$, $23^\circ$ and $33^\circ$ at $z=0$, $1$ and $2$, respectively. Analyzing $N$ independent galaxies reduces the uncertainty on an isotropic cosmic birefringence signal to $\sigma_{\alpha,{\rm eff}}/\sqrt{N}$, providing competitive sensitivity once large samples are available. As accurate observation-driven models are not yet available for cosmological galaxy samples, our results motivate pilot observations to empirically characterize polarization-shape alignment, and can facilitate forecasts of cosmology and fundamental physics applications that exploit this effect.
}

\maketitle

\section{Introduction} 
\label{sec:intro}

Synchrotron radiation is the electromagnetic radiation emitted when cosmic-ray electrons, accelerated by stellar winds and supernova explosions, spiral around interstellar magnetic fields. In star-forming galaxies such as the Milky Way, this diffuse component provides the dominant contribution to the $1$–$10$ GHz radio continuum, as long as the central active galactic nucleus (AGN) remains weak or absent. In the cosmological volume, such ``normal'' galaxies~\citep{Condon1992ARAAreview} have a higher abundance than the Seyfert galaxies, and much more so than the AGN-powered radio-loud galaxies. They populate the faint end of the radio luminosity function (RLF)~\citep{MauchSadler2007RadioSourcesIn6dFGS} and require the deepest radio continuum surveys to be studied (e.g.~\cite{Smolcic2017VLACOSMOS3GHzLargeProject, Hale2025MIGHTEEsurveyDR1}).

The polarization direction of this ISM emission, set locally by the magnetic field, exhibits an ordered spatial pattern across the disk since the magnetic fields have an ordered component tracing spiral arms. When the disk appears inclined, radio continuum emission integrated over the galaxy has a net polarization closely aligned with the apparent short axis of the disk. Stil {\textit{et al.}}~\cite{stil2009integrated} first observed this polarization-shape alignment in nearby spiral and barred galaxies using radio polarimetry data at $4.8\,$GHz. Spiral galaxies are found to show the tightest alignment with a root-mean-square (RMS) misalignment angle $\sim 20^\circ$--$30^\circ$. Barred galaxies show this effect to a lesser degree. The integrated polarization fraction is low for nearly face-on galaxies as a result of geometric cancellation, and is also low for edge-on galaxies due to Faraday depolarization through the gas disk.

These results have since prompted keen interest among cosmologists in exploiting polar\-ization-shape alignment for cosmology and fundamental physics measurements. For one example, foreground weak lensing distorts the observed galaxy shape, but does not change the integrated polarization direction. In pursuit of this idea, methods have been proposed to measure cosmic shear, in which the galaxy intrinsic shape is to be calibrated with radio polarimetry so that the confounding effect of intrinsic shape alignment can be mitigated~\cite{BrownBattye2011weaklensingIA, Camera2017SKAIII}. Ref.~\cite{Jarvis2016MIGHTEEsurvey} considered exploiting this effect to augment cosmic shear measurements using polarization information from the MIGHTEE survey.

Conversely, calibrating intrinsic polarization direction to the galaxy apparent shape enables extraction of polarization rotation along the line of sight, which is a smoking gun for axion fields that couple to electromagnetism through the Chern-Simons interaction~\cite{Carroll:1989vb, Harari:1992ea, Fedderke:2019ajk, Agrawal:2019lkr}. Recently, Ref.~\cite{yin2025new} proposes a new method to measure this cosmic birefringence by leveraging and combining integrated radio polarimetry data and galaxy shape data. The authors develop a minimal-variance quadratic estimator for the angular power spectrum of the polarization rotation angle, and estimated the number of usable galaxies assuming reference Square Kilometre Array (SKA) radio continuum surveys combined with major photometric shape catalogs. This observable complements detection methods based on cosmic microwave background (CMB) polarization anisotropies~\cite{Kamionkowski2009, Gluscevic2009, Yadav2009, Yin:2021kmx} in that available galaxies probe birefringence sources during the most recent history of cosmic expansion at $z= 0$--$2$. When supplemented with spectroscopic redshifts, the method can allow redshift-resolved measurements~\cite{Naokawa:2025shr}, in addition to CMB-based tomographic methods already studied~\cite{Sherwin:2021vgb, Lee:2022udm, Nakatsuka:2022epj, Namikawa:2023zux}. Ref.~\cite{Naokawa:2025shr} proposes the same method using radio galaxies to measure uniform cosmic birefringence induced by a slowly-rolling ultralight axion field, in order to test a low-$z$ physical origin of the reported uniform polarization rotation angle in Planck~\citep{Minami:2020odp, Eskilt:2022cff, Diego-Palazuelos:2022dsq, Cosmoglobe:2023pgf} and ACT~\citep{ACT:2025fju} data.

However, it is unknown whether galaxies at cosmological distances exhibit polarization-shape alignment similar to what are found by Ref.~\cite{stil2009integrated}. After all, intrinsic properties of star-forming galaxies evolve significantly with redshift. The geometric regularity of the disk structure and the ISM magnetic fields can vary with both redshift and halo mass. The sample of Ref.~\cite{stil2009integrated} only include a few dozen galaxies within $40\,$Mpc, 13 of which are classified as spirals. While the alignment is expected over a range of radio frequencies, the extent to which internal Faraday depolarization diminishes the effect toward lower frequencies $\nu \lesssim 4.8\,$GHz is unknown. To our knowledge, there have not been published observations that generalize the findings of Ref.~\cite{stil2009integrated} to a cosmological sample of galaxies. Without answers to these questions, the potential of the aforementioned cosmological applications is hard to assess~\citep{Harrison2016SKAweaklensingI}.

\begin{figure*}[htbp]
\includegraphics[width=\linewidth,scale=1.00]{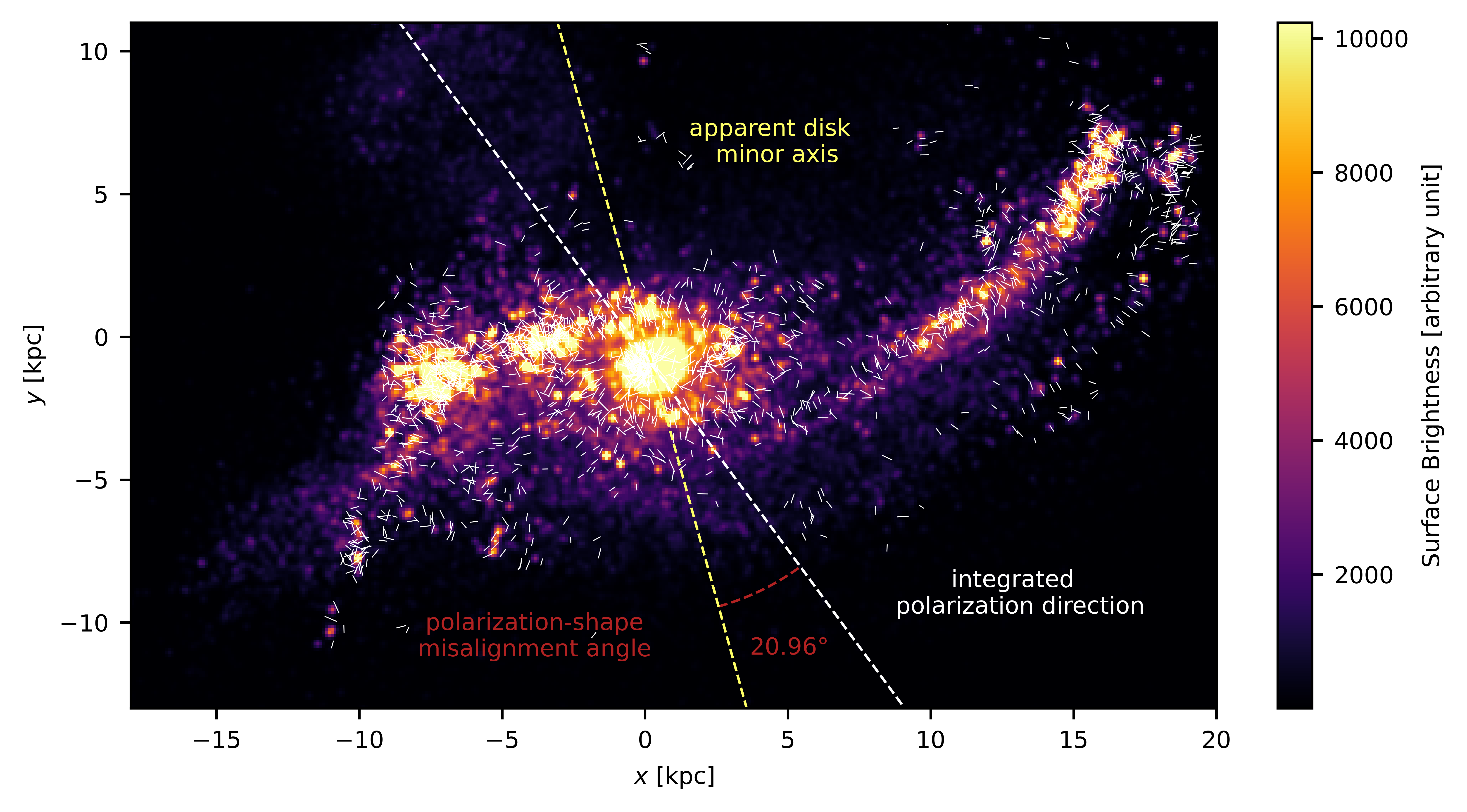}
\caption{Illustration of the polarization-shape alignment effect. We show the V-band stellar surface brightness profile of a simulated $z=0$ galaxy (\texttt{IllustrisTNG-50} Subhalo ID: 19) with its inclined disk. The yellow dashed line marks the apparent semi-minor axis of the galaxy, which is determined from optical photometric moments. White segments represent synchrotron emission at $4.8\,$GHz near major sites of star formation. The length of the segment is proportional to the local intensity of the polarized emission $\sqrt{Q^2+U^2}$, and the orientation indicates the direction of polarization. For clarity, only 1 in 20 of the segments is shown. The white dashed line marks the polarization direction of synchrotron emission integrated over the entire galaxy. This galaxy shows a misalignment angle $\Delta\theta=20.96^\circ$ between polarization and shape.}
\label{fig:example_galaxy}
\end{figure*}

In this paper, we find tentative answers to these questions by studying simulated galaxies from the \texttt{IllustrisTNG} project~\cite{Weinberger2017TNGmodel, Phillepich2018IllustrisTNGmodel, Nelson2019IllustrisTNG50}. To this end, we compute the polarized synchrotron emission powered by star formation, and measure the optical galaxy shapes, for a large number of galaxies selected from \texttt{IllustrisTNG} simulation boxes. \reffig{example_galaxy} exemplifies polarization-shape alignment measured for one simulated galaxy. We aim to derive statistics of the misalignment angle and quantify variation with disk inclination, radio frequency and redshift. Ultimately, our results should be compared to data from cosmological radio continuum surveys. This work will sharpen our understanding of this polarization-shape alignment phenomenon, and will pave the way for exploiting this phenomenon for cosmology and fundamental physics research.


The remainder of this paper is organized as follows. In \refsec{methods}, we discuss galaxy selection from \texttt{IllustrisTNG50} and describe methods we use for computing the apparent galaxy shape and the polarized synchrotron emission. These procedures enable us to derive statistics for the polarization-shape misalignment angle for simulated galaxies. In \refsec{results}, we present key results from this analysis, which include the radio luminosity function, distribution of integrated polarization fraction, distribution of polarization-shape misalignment angle, and Fisher information for estimating a uniform external polarization rotation angle. We will discuss our findings in \refsec{discuss}, before we make concluding remarks in \refsec{concl}. Technical details are relegated to several Appendices for interested readers: \refapp{galaxy_shape} on the method to measure galaxy shapes, \refapp{sync_emis} on the calculation of polarized synchrotron emission, \refapp{ne} on the method to extract ISM electron number density from \texttt{IllustrisTNG50} simulation data, and \refapp{subgridBfields} on a sub-grid treatment of randomly oriented ISM magnetic fields unresolved in the simulation.

\section{Methods} 
\label{sec:methods}

In this Section, we first discuss the public galaxy formation simulation data we use and our galaxy sample selection criteria. We then outline our methodology for modeling diffuse synchrotron emission powered by star formation with full polarization information. We also describe how we characterize polarization-shape (mis-)alignment using simulation data.

\subsection{Simulations and galaxy selection}
\label{sec:theillustrisproject}

We analyze galaxies simulated in the \texttt{IllustrisTNG50} suite. This project is the third and final volume of the IllustrisTNG cosmological magnetohydrodynamical (MHD) simulation series~\cite{Nelson2019IllustrisTNG50}. It simulates a cosmological cube of side length $50\,$cMpc, which is sampled by $2160^3$ cells with an average spatial resolution of $70$--$140\,$pc in star-forming regions~\cite{2019MNRAS.490.3234N, 2019MNRAS.490.3196P}. We choose this \texttt{IllustrisTNG} suite with the highest spatial resolution in order to resolve as much as possible the small-scale details of star-formation feedback and interstellar magnetohydrodynamics, as we will have to sum over cells to find integrated polarized radio emissions. \texttt{IllustrisTNG50} adopts a $\Lambda$CDM cosmology in concordance with the Planck 2015 cosmological results~\cite{Planck:2015fie}, with parameters $\Omega_{\Lambda,0} = 0.6911$, $\Omega_{\mathrm{m},0} = 0.3089$, $\Omega_{\mathrm{b},0} = 0.0486$, $\sigma_8 = 0.8159$, $n_{\mathrm{s}} = 0.9667$, and $h = 0.6774$. For the purposes of this study, these are indistinguishable from more recent cosmology model fits.

We are primarily interested in Milky Way (MW)-like galaxies with a disk. We therefore select host dark matter halos with masses in the range $M_{\rm halo} = 10^{11}$--$10^{13}\,M_\odot$. Larger dark matter halos tend to host massive elliptical galaxies that lack both star formation activity and a disk structure. On the other hand, smaller dark matter halos often host immature galaxies with lower stellar masses and more irregular morphologies. Interested in radio continuum emission powered by star formation feedback, we additionally require the selected galaxies to have a significant star formation rate (SFR) in the range ${\rm SFR}=0.2$--$10\,M_\odot\,{\rm yr}^{-1}$, in order to exclude both quiescent galaxies and rare starburst galaxies with exceptionally high SFRs. This second criterion lead to selected galaxies that have similar properties to the MW. In \reftab{galaxy_number_density}, we list the number of selected galaxies in six different redshift snapshots. In light of cosmological applications of polarization-shape alignment, these redshifts cover the redshift range $z=0$--$1.5$ feasible with forthcoming radio continuum surveys~\citep{yin2025new}. The upper SFR cutoff leaves out only an insignificant fraction of galaxies by number. At low redshifts, our selected galaxies indeed have a comoving number density $\sim 10^{-2}\,{\rm cMpc}^{-3}$ similar to dark matter halos hosting MW-like galaxies~\citep{Lukic:2007fc}.


\begin{table}[htbp]
    \caption{Total number of galaxies $N_{\rm gal}$ selected in six different IllustrisTNG50 redshift snapshots and the corresponding comoving number densities $n_{\rm gal}$.}
    \centering
    \begin{tabular}{@{\extracolsep{4pt}}ccc}
        \hline\hline
        Redshift $z$ & $N_{\rm gal}$ & $n_{\rm gal}$ $[{\rm cMpc}^{-3}]$ \\
        \hline
        0   & 1531 & $1.22 \times 10^{-2}$ \\
        0.3 & 1733 & $1.39 \times 10^{-2}$ \\
        0.7 & 1684 & $1.35 \times 10^{-2}$ \\
        1.0 & 1604 & $1.28 \times 10^{-2}$ \\
        1.5 & 1410 & $1.13 \times 10^{-2}$ \\
        2.0 & 1157 & $0.93 \times 10^{-2}$ \\
        \hline\hline
    \end{tabular}
    \label{tab:galaxy_number_density}
\end{table}

\subsection{Apparent galaxy shape}

The projected disk shape on the sky needs to be measured in order to characterize polarization-shape (mis)alignment. Shape measurements are often carried out for many existing and forthcoming imaging surveys to enable weak lensing measurements. In fact, in Ref.~\cite{stil2009integrated} the integrated radio polarization orientation is compared to that of the disk ellipse extracted from optical images. 

We calculate the apparent shape of the simulated galaxies based on spatially resolved starlight surface brightness provided in \texttt{IllustrisTNG50}. We assume a simple ellipse model in which the projected galaxy shape is characterized by only three parameters: the semi-major axis $a$, the semi-minor axis $b$, and the ellipse position angle $\theta_e$. These quantities are computed in three photometric bands, U, V, and i, using the \texttt{GFM\_StellarPhotometrics} field from the \texttt{IllustrisTNG} simulation data release~\cite{nelson2021illustristngsimulationspublicdata}. These quantities are computed based on photometric moments, which are direct observables independent of galaxy shape models. See Appendix~\ref{app:galaxy_shape} for technical details.

\reffig{inclination_dist} compares the measured inclination distribution to the theoretical isotropic distribution. The agreement is fair at low to intermediate inclination values, but there appears to be a dearth of nearly edge-on galaxies. This likely results from the neglect of a prominent bulge, which inflates the $b/a$ ratio. More quantitatively, if the disk is very close to edge-on so that $b/a$ is dominated by the presence of a bulge (or a bar), then $b/a$ is on the order of a small parameter $\varepsilon$, which is the ratio between the bulge size and the disk size. The rapid drop in count beyond $i \approx 70^\circ$ in \reffig{inclination_dist} suggests a typical $\varepsilon\approx \cos70^\circ =0.34$. As long as it is not reflection symmetric with respect to the disk, the presence of the bulge should also affect determination of the apparent disk major/minor axes. However, according to \refEq{theta_e_def} the orientation of the axes as measured using photometric moments is perturbed by an amount up to $\sim \varepsilon^2$ in radians, which is only $\sim 7^\circ$. Given our interest here in quantifying (mis)alignment between the integrated polarization direction and the photometrically inferred shape orientation, this $\sim 7^\circ$ perturbation will be mostly a subdominant effect compared to the size of the intrinsic misalignment, as is later shown in \reffig{delta_fit} for an example. Therefore, while photometric measurement leads to significantly biased inclination determination for high inclination galaxies, we expect that degrading of the apparent polarization-shape alignment will only be modest.

In principle, inclination can be more correctly determined by the direction of the total angular momentum vector of the galaxy's stellar mass, which can well reproduce the isotropic distribution~\cite{Kovacs_2024} unlike in \reffig{inclination_dist}. Unfortunately, the stellar angular momentum vector is not directly measurable in galaxy imaging surveys. This problem would have to be overcome with spatially resolved spectroscopic mapping of the disk, namely the kinematic weak lensing method (e.g. \cite{Blain2002CosmicShearWithSpectralImaging, Huff2013CosmicShearWithoutShapeNoise}). For our purposes, the method based on photometric moments is more practical. In the remainder of this paper, we shall bin simulated galaxies based on the apparent inclination measured using photometric moments. \reffig{inclination_dist} also shows that the measured inclination is consistent across U, V and i bands. For the rest of this paper, we will use shapes measured in the V band.

\begin{figure}[t]
\includegraphics[width=0.7\textwidth]{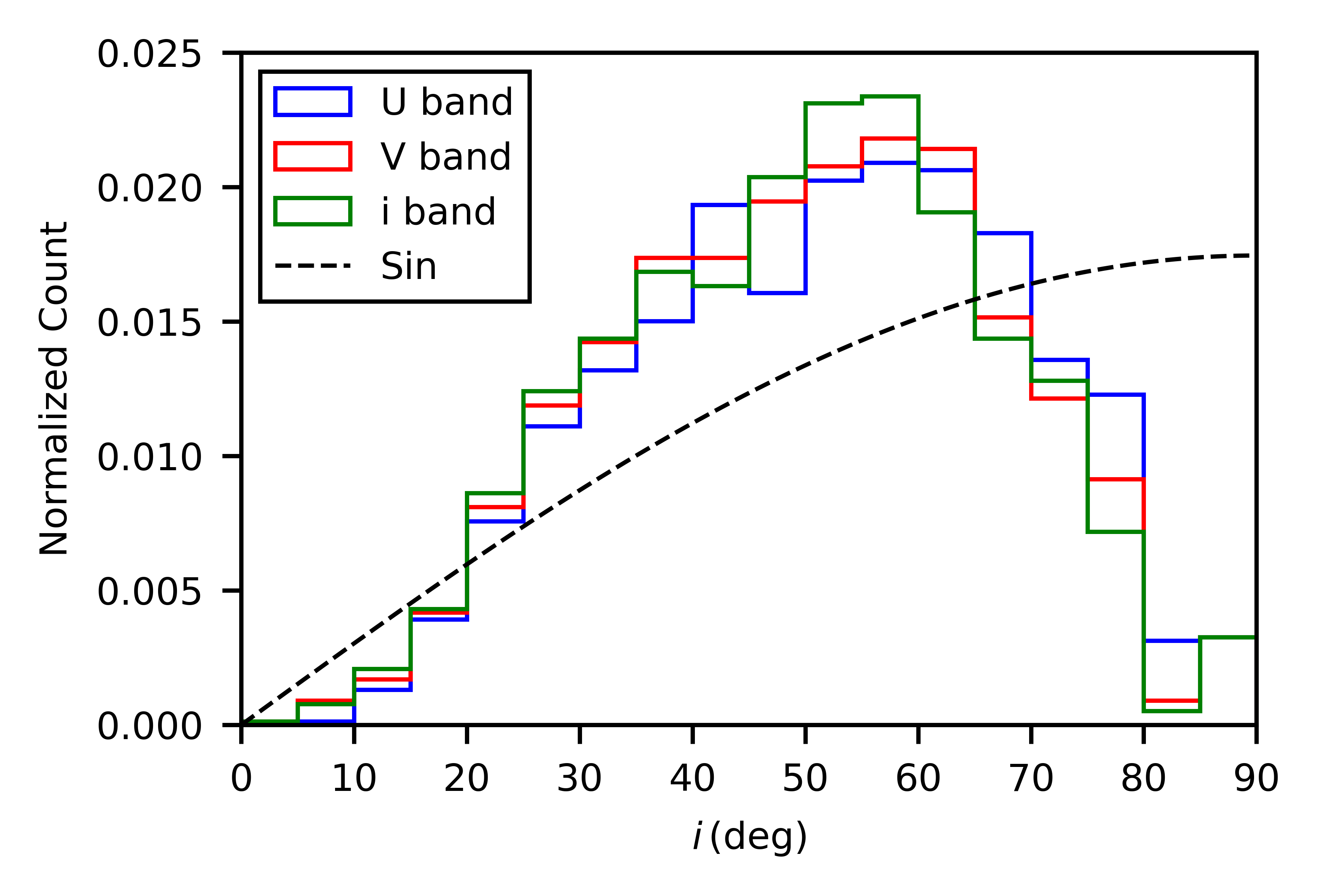}
\centering
\caption{Distribution of inclination $i$ inferred from photometric moments (colored histograms) compared to the theoretical isotropic distribution (black histogram). Inclinations are estimated from the axial ratio $b/a$ (see Appendix~\ref{app:galaxy_shape}), assuming all galaxies are thin disks without bulges. The discrepancy with the isotropic distribution at very high $i$ may arise from neglecting the bulge in our thin-disk modeling of galaxy shapes.
\label{fig:inclination_dist}}
\end{figure}

\subsection{Synchrotron emission}

Star formation (SF) and active galactic nuclei (AGN) are two primary energy sources of ISM synchrotron emission at one to several GHz~\citep{Hansen2024SHARKmodel}. While the most radio-loud galaxies are typically powered by AGNs, star formation dominates the diffuse radio emission for a larger fraction of the galaxy population with low radio luminosities. This star-formation dominated galaxy population is the focus of our study.

Non-thermal electrons are not part of the base physics treated in \texttt{IllustrisTNG50}. We thus have to predict synchrotron emission using other information available from the simulation. Given the magnetic field in each simulation cell, we can calculate the polarized emissivity from that cell assuming optically thin conditions. Useful calculation details are collected in Appendix~\ref{app:sync_emis}. 

The normalization of synchrotron luminosity in each cell is set to be proportional to the star-formation rate in that cell. We do not relate synchrotron luminosity to local ISM magnetic energy. The constant of proportionality between synchrotron luminosity and star-formation rate does not affect the polarization direction of the integrated emission, provided that this constant applies to all cells in a given galaxy. Normalizing synchrotron emission to star-formation rate is motivated by the observed spatial similarity between synchrotron emission and mid-infrared emission~\citep{Murphy2006FIRradioCorrelation, Murphy2006FIRradioCorrSpitzer}, since both are energetically sourced by star formation. In a more detailed picture, relativistic electrons diffuse $\sim 0.1$--$1\,$kpc away from the acceleration site before they cool down in $\sim 100\,$Myr, so synchrotron surface brightness is found to only partially correlate with star formation~\cite{Murphy2006FIRradioCorrelation, Tabatabaei2013RadioFIRCorrNGC6946}. Given the finite cell sizes $\sim 100\,$pc in \texttt{IllustrisTNG50}, we adopt the local, cell-by-cell normalization prescription for simplicity.

At frequencies $\nu=1$--$10\,$GHz, we neglect free-free absorption and synchrotron self-absorption, which are important only in exceptionally dense or highly magnetized ionized regions. Later, we shall account for internal Faraday rotation as radio waves emitted within each cell propagate through the ISM of the host galaxy toward the observer.

\begin{figure}[htbp]
\includegraphics[width=0.7\textwidth]{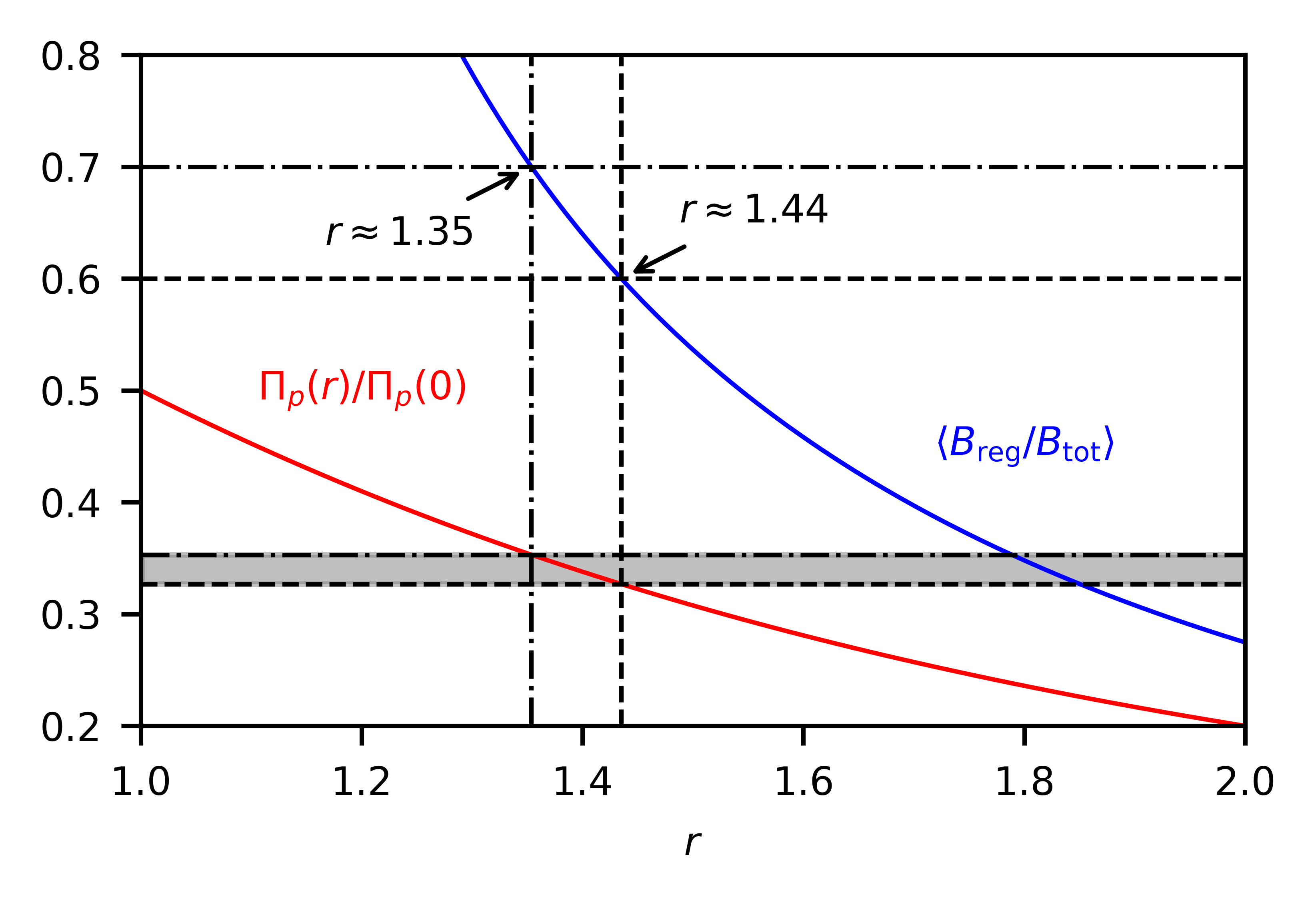}
\centering
\caption{Modification of polarization fraction due to unresolved random magnetic field components. Here $r$ is the ratio between the random component and the regular component of the magnetic field. Requiring $\langle B_{\rm reg}/B_{\rm tot} \rangle$ to be $0.6$--$0.7$ from Milky-Way ISM observations~\citep{Beck2004role}, the ratio $r=1.35$--$1.44$. This leads to a constant reduction factor $\Pi_{p}(r)/\Pi_{p}(0)$ ranging from $0.327$ to $0.353$ for polarization fraction from all simulation cells, where $\Pi_{p}(r)$ is the polarization fraction in the presence of a random magnetic field component, and $r \equiv B_{\rm rand} / B_{\rm reg}$ is the ratio of the random to regular magnetic field strengths.
\label{fig:b_field_subgrid}}
\end{figure}

Integration of polarized synchrotron emission assuming a uniform magnetic field within each cell turns out to over-predict the polarization fraction of the total signal. This is almost certainly due to neglect of small-scale randomly-oriented ISM magnetic fields that are under-resolved in \texttt{IllustrisTNG50}. The presence of such random fields suppresses the overall polarization from each cell because Stokes parameters from different parts within a single cell do not perfectly align. In Appendix~\ref{app:subgridBfields}, we employ a simple analytic sub-grid treatment to account for this frequency-independent depolarization. Assuming a typical relative strength between the regular and random components as found for ISM magnetic fields in the MW~\citep{Beck2004role}, we multiply the polarized Stokes parameters from all cells by a constant factor $0.327$ (see Figure~\ref{fig:b_field_subgrid}).

\subsection{Internal Faraday depolarization}

Magnetized ionized gas rotates the polarization plane of radio waves by an angle $\Delta \theta_{\rm RM} = {\rm RM}\,\lambda^2$, where the rotation measure RM, in units of ${\rm rad}\,{\rm m}^{-2}$, is given by the line of sight integral
\begin{align}
    {\rm RM} = 0.81\,\int\,n_e\,B_{\parallel}\,\rmd l.
\end{align}
Here $n_e$ is the electron number density in $\mathrm{cm}^{-3}$, $B_\parallel$ is the magnetic field component along the line of sight in $\mu\mathrm{G}$, and $l$ is the line-of-sight path length in $\mathrm{pc}$~\cite{jung2023samplingfaradayrotationsky, Kovacs_2024}.

For a distant source galaxy, Faraday rotation accrued through the ISM in the MW is expected to be coherent across the source galaxy and thus does not change the integrated polarization fraction. This may be (partially) corrected for if multi-frequency data are available. On the contrary, Faraday rotation within the source galaxy is inhomogeneous between different cells, and hence reduces the integrated polarization fraction. This is the internal Faraday depolarization effect, which is independent of the modification discussed in the previous section, where we addressed unresolved magnetic fields within individual cells. Such internal depolarization is stronger at lower frequencies, and significantly degrades polarization-shape alignment as shown in \reffig{sig_alpha_eff}.

To account for internal Faraday rotation, we first define a radius, $r_{{\rm SF},99}$, which encloses $99\%$ of the star-forming cells. Then we choose an integration step length $50\,{\rm pc}$ for a balance between accuracy and computational time cost. Rotation angle $\Delta \theta_{\mathrm{RM}}$ accumulated along the line of sight is computed for each cell and is applied to the Stokes parameters~\citep{Kovacs_2024}. Appendix~\ref{app:ne} explains how the electron number density $n_e$, which is needed for calculating Faraday rotation, is derived from simulation data.

\subsection{Polarization-shape misalignment angle}

The polarization direction of the integrated synchrotron emission closely aligns with the apparent minor axis of an inclined disk galaxy~\citep{stil2009integrated, yin2025new}. We therefore define the polarization-shape misalignment angle as
\begin{align}
\label{eq:Delta_theta}
    \Delta \theta = \theta_{p} - \theta_{e} + \pi / 2,
\end{align}
which lies in the range $(-\pi / 2, \pi / 2)$. For our selected galaxies, the polarization position angle $\theta_p$ is computed from the spatially integrated Stokes parameters, and the shape position angle $\theta_e$ is computed from stellar light moments. This is exemplified for one simulated galaxy in \reffig{example_galaxy}. In this way, we collect statistics of $\Delta\theta$ for the simulated galaxies.

\begin{figure*}[htbp]
\includegraphics[width=\textwidth]{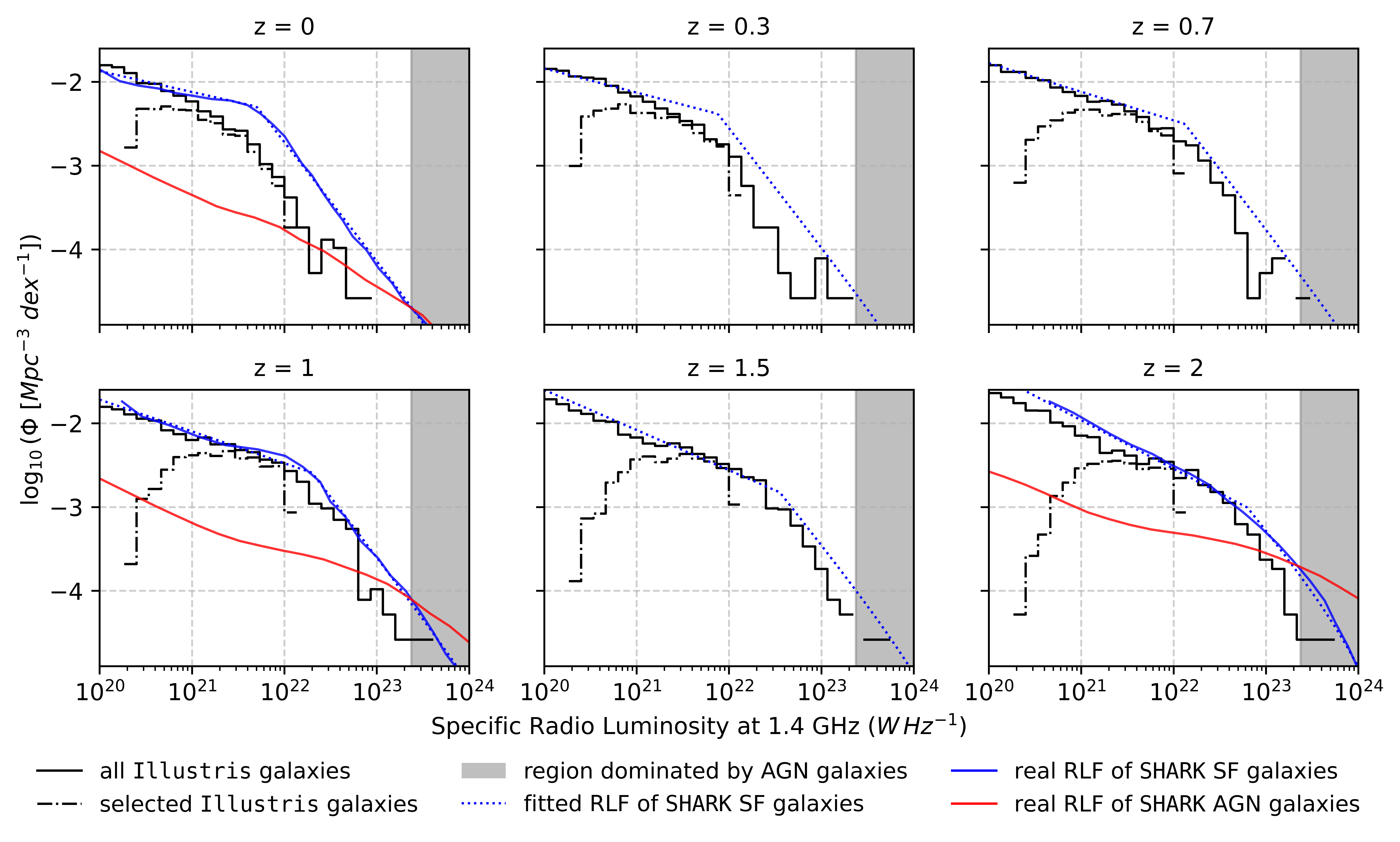}
\caption{Radio galaxy luminosity function (RLF) at $1.4\,$GHz for each of the six \texttt{Illustris} redshift snapshots we analyze. In each panel, the dash-dotted histograms show our selected \texttt{IllustrisTNG50} star-forming galaxies with a halo mass $M_{\rm halo}=10^{11}$--$10^{13}\,M_\odot$ and star formation rate $0.2$--$10\,M_\odot\,{\rm yr}^{-1}$, while the solid histogram includes all \texttt{IllustrisTNG50} galaxies. The luminosities include only star formation contribution for the black (solid and dash-dotted) histograms. All luminosities are estimated using the analytic recipes of \cite{Hansen2024SHARKmodel}. For comparison, the blue curve shows the star-formation RLF taken from \cite{Hansen2024SHARKmodel}, and should be compared to the solid black histogram. The red curve shows the AGN RLF from the same work. These results are available and shown for $z=0$, $1$ and $2$. The blue dotted curve is an analytic fit to the RLF, consisting of a power-law at low luminosities and a Schechter function at high luminosities.
The grey shaded region indicates the luminosity range where AGN-dominated radio galaxies are expected to outnumber those dominated by star formation~\citep{Hansen2024SHARKmodel}.
The majority of the detectable radio galaxies at $z<1.5$ will be star-formation dominated. 
\label{fig:radio_LF}}
\end{figure*}

\section{Results}
\label{sec:results}

In this section, we present and discuss results based on our calculations of the integrated synchrotron emission, its polarization, and the polarization-shape misalignment angle.

\subsection{Radio luminosity function}

By surveying synchrotron luminosities of our selected galaxies and dividing by the comoving volume of the simulation box, we derive the radio luminosity functions (RLF), which is shown in \reffig{radio_LF}. We have not specified normalization of the synchrotron emissivity when calculating the integrated polarization fraction. To normalize the radio emission, here we use the analytic prescription of Ref.~\citep{Hansen2024SHARKmodel}, which provides separate models for the radio emission powered by star formation and by AGN activity. Here we only calculate the former contribution, which should be appropriate for ``normal'' galaxies with very weak or negligible AGN lobe and/or core emissions~\citep{Condon1992ARAAreview}. In \texttt{IllustrisTNG50}, many galaxies are not seeded with a supermassive blackhole (SMBH) as their dark matter halos do not reach the threshold mass. Moreover, simulation resolution is insufficient for accurately resolving magnetohydrodynmaic properties in the galactic center region. We therefore refrain from calculating the AGN emission using simulation data, which would likely be unreliable. We will comment on the neglected AGN contribution and its impact on the polarization-shape alignment statistics later in \refsec{AGNemission}.

\reffig{radio_LF} shows that when all \texttt{IllustrisTNG50} galaxies are included, the derived RLF closely matches the observation-calibrated analytic model of \citep{Hansen2024SHARKmodel} in the low luminosity range $\log L_{\nu, 1.4}[{\rm W}\,{\rm Hz}^{-1}] < 21$, but underpredicts the source number for $21 < \log L_{\nu, 1.4}[{\rm W}\,{\rm Hz}^{-1}] < 23$ by a factor of a few. Among those, our selected galaxies make up the bulk in the luminosity range $21 < \log L_{\nu, 1.4}[{\rm W}\,{\rm Hz}^{-1}] < 22$. Qualitatively similar conclusions can be drawn at the other redshifts we study.

\reffig{radio_LF} shows that the bulk of our selected galaxies lie well below the luminosity threshold where AGNs start to dominate the RLF, according to \cite{Hansen2024SHARKmodel}. The continuum 5$\sigma$ detection limits (which is required for meaningful polarization measurements) around $1.4\,$GHz in three benchmark SKA-MID continuum surveys considered in \cite{yin2025new}, namely the \texttt{Wide}, \texttt{Deep} and \texttt{Ultradeep} reference surveys with Band 2 $1\sigma$ depths $\sigma_I=1.0,\,0.2,\,0.05\,\mu{\rm Jy}$, respectively, lie in the low luminosity range where star-formation power sources are more abundant than AGN powered sources, at least for $z \lesssim 1$. Most SKA detectable galaxies usable for the polarization-shape alignment probe will be from $z\lesssim 1.5$~\citep{yin2025new}. We therefore believe that the polarization-shape (mis)alignment statistics we will present should be applicable to a significant fraction of the galaxies detectable as radio continuum sources in those SKA surveys.

\subsection{Polarization fraction}

\begin{figure*}[htbp]
\includegraphics[width=\textwidth]{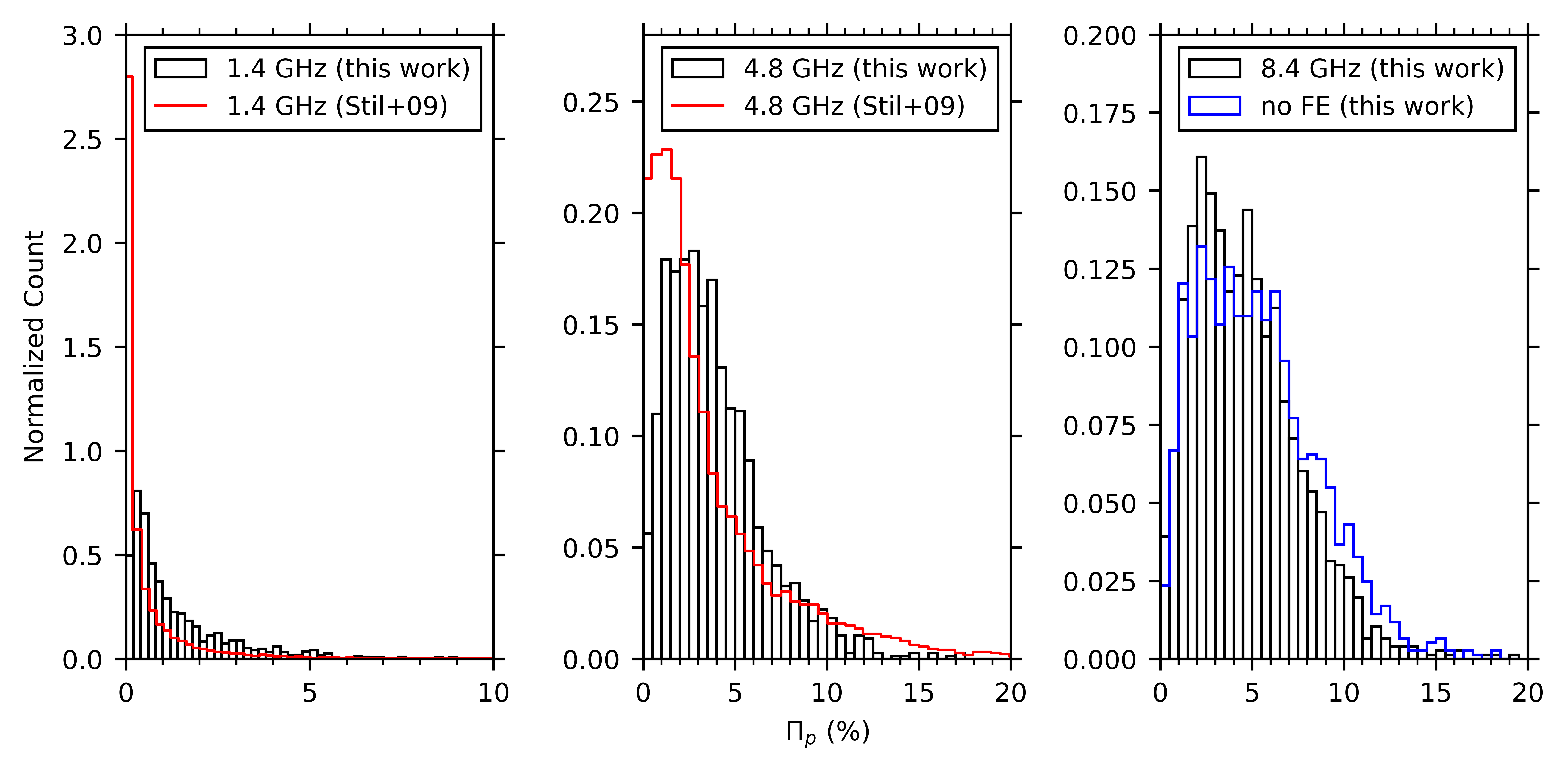}
\caption{Computed polarization fraction $\Pi_{p}$ for our selected \texttt{IllustrisTNG50} galaxies at $z=0$, at three different frequencies and with internal Faraday depolarization (FE) accounted for. The distributions show fair agreement with independent model predictions from \cite{stil2009integrated} (red histograms). Internal Faraday depolarization significantly reduces the polarization fraction at $1.4\,$GHz and $4.8\,$GHz, but is unimportant at $8.4\,$GHz. 
\label{fig:deg_of_pol}}
\end{figure*}

Panels of \reffig{deg_of_pol} show the distribution of integrated polarization fraction $\Pi_{p}$ for our selected galaxies at $z=0$, computed for $\nu=1.4,\,4.8,\,8.4\,$GHz, respectively. The effects of sub-grid magnetic fields and internal Faraday depolarization are accounted for. Internal Faraday depolarization appears insignificant at $8.4\,$GHz, but significantly reduces $\Pi_{p}$ at $4.8\,$GHz, and much more so at $1.4\,$GHz. The histograms are largely in agreement with the predictions of the semi-analytic toy model developed in Ref.~\cite{stil2009integrated}. There are no published large integrated polarization fraction statistics which we can compare our results to.

\subsection{Cosmic birefringence estimation}

In the presence of cosmic birefringence induced by an axion-like scalar field along the line of sight, the observed polarization direction rotates by a small angle $\alpha$ relative to the intrinsic direction, while the apparent galaxy shape is unaffected. This induces a subtle change in the statistics of polarization-shape alignment, which can be exploited to measure $\alpha$. In particular, the misalignment angle defined in \refEq{Delta_theta} shifts, $\Delta\theta \rightarrow \Delta\theta + \alpha$.

Ref.~\cite{yin2025new} devised an unbiased estimator for the rotation angle,
\begin{equation}
\label{eq:alpha_minvar}
    \hat\alpha = \frac{\sigma_Y^2\,\overline X\,Y + (\sigma_X^2 - \sigma_Y^2)\,X\,Y}{2\left[(\sigma_X^2 - \sigma_Y^2)^2 + \sigma_X^2\,\overline X^2\right]},
\end{equation}
where $X$ and $Y$ are spin-0 products of radio Stokes parameters and ellipticity, with even and odd parity, respectively (see \cite{yin2025new} for definitions). \refEq{alpha_minvar} is an unbiased quadratic estimator with a minimized variance under the null hypothesis,
\begin{equation}
    \langle{\hat\alpha^2}\rangle = \frac{\sigma_X^2\,\sigma_Y^2}{4\left[(\sigma_X^2 - \sigma_Y^2)^2 + \sigma_X^2\,\overline X^2\right]},
    \label{eq:var}
\end{equation}
provided that $X$ has a Gaussian distribution of mean $\overline{X}$ and variance $\sigma^2_X$, and $Y$ has a Gaussian distribution of zero mean and variance $\sigma^2_Y$.

\begin{figure}[htbp]
\includegraphics[width=0.6\textwidth]{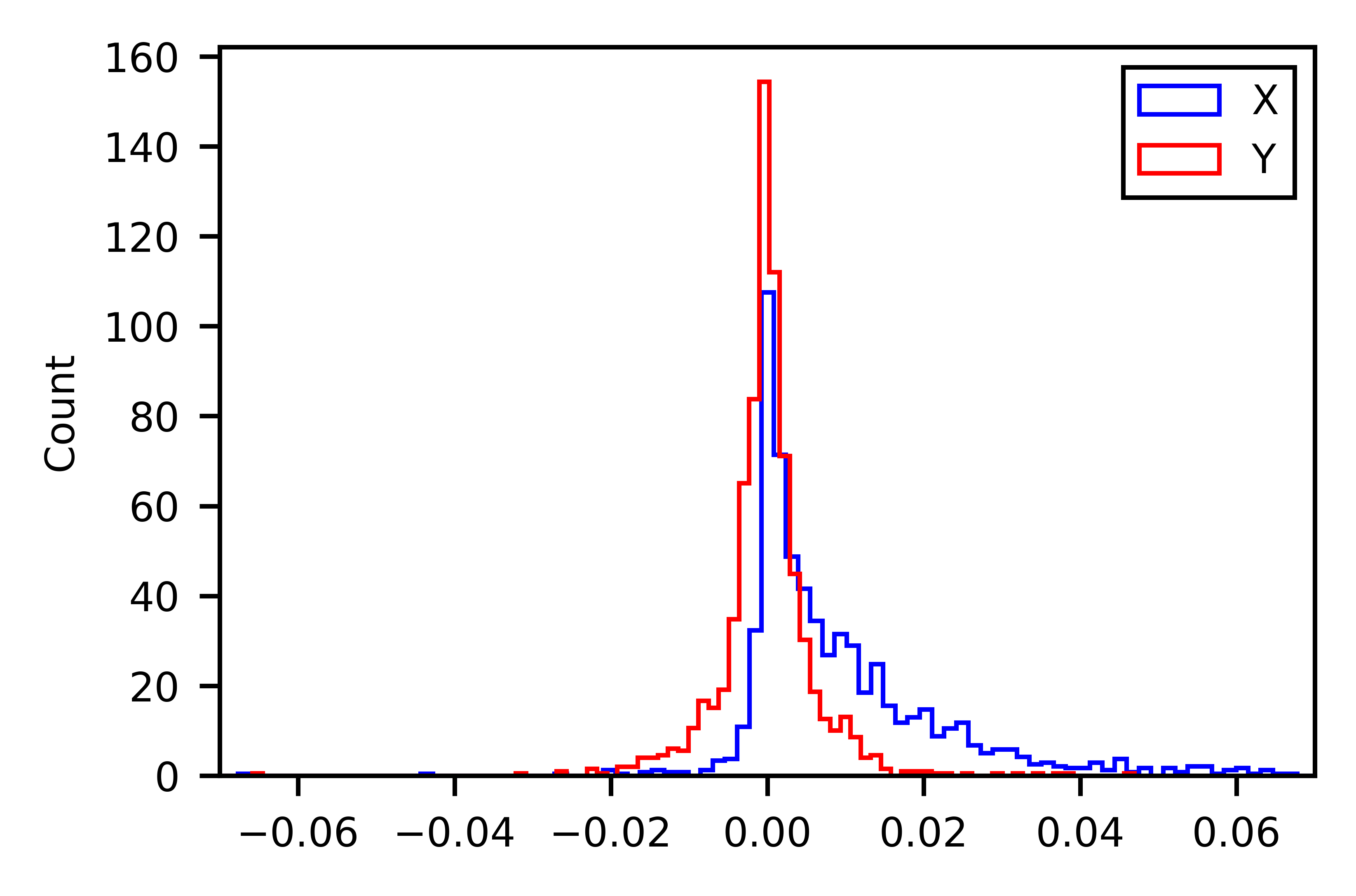}
\centering
\caption{The spin-0 quantities $X$ and $Y$ as defined in Ref.~\cite{yin2025new} have non-Gaussian distributions. Example statistics for selected \texttt{IllustrisTNG50} galaxies are shown here for radio polarization measured at $4.8\,$GHz in $z = 0$ galaxies.}
\label{fig:non_gaussian_x_y}
\end{figure}

With the small galaxy sample of Ref.~\cite{stil2009integrated}, it could not be reliably determined whether those assumptions about the distribution of $X$ and $Y$ are valid. Our measurements of the \texttt{IllustrisTNG50} galaxies indicate that $X$ and $Y$ have non-Gaussian and skewed distributions, especially when all disk inclinations are included (see \reffig{non_gaussian_x_y} for an example). This renders it unreliable to forecast sensitivity to measuring $\alpha$ via \refEq{var}, as the answer turns out to be sensitive to outliers in $X$ and $Y$.

To account for non-Gaussianity in the statistics, we instead evaluate the Cram\'{e}r-Rao bound on estimation of $\alpha$ using Fisher information theory, which is applicable to any general unbiased estimator for $\alpha$. Consider the joint probability distribution function $P(\Delta\theta,\,\Pi_p)$ that characterizes the intrinsic property of a given galaxy population, where $\Delta\theta$ is the misalignment angle as in \refEq{Delta_theta} and $\Pi_p=\sqrt{Q^2+U^2}/I$ is the polarization fraction. Under external polarization rotation by a constant angle $\alpha$, $\Delta\theta \rightarrow \Delta\theta+\alpha$, but $\Pi_p$ does not change. The Fisher information in estimating $\alpha$, calculated under the null hypothesis $\alpha=0$, is
\begin{align}
\label{eq:Fisher_F}
    F & = \iint\rmd\Delta\theta\,\rmd \Pi_p\,P(\Delta\theta,\,\Pi_p)\,\left(\frac{\partial \ln P(\Delta\theta,\,\Pi_p)}{\partial\Delta\theta}\right)^2 \nonumber\\
    & \geqslant \int\rmd\Delta\theta\,P(\Delta\theta)\,\left(\frac{\partial \ln P(\Delta\theta)}{\partial\Delta\theta}\right)^2,
\end{align}
where $P(\Delta\theta)=\int\rmd \Pi_p\,P(\Delta\theta,\,\Pi_p)$ is the marginalized distribution for the misalignment angle. The second line follows from the Cauchy-Schwarz inequality, which is saturated only if $\Delta\theta$ and $\Pi_p$ are statistically uncorrelated, i.e. $P(\Delta\theta,\,\Pi_p)=P(\Delta\theta)\,P(\Pi_p)$. As we will discuss later, this condition is not true due to significant dependence of misalignment on inclination.

A simple treatment is to disregard $\Pi_p$ and calculate Fisher information in the marginalized $P(\Delta\theta)$. This distribution is poorly fit by a Gaussian function, since it exhibits a narrow peak at small $\Delta\theta$ values and extended wings at large $\Delta\theta$. 
Instead, we consider the following one-parameter distribution
\begin{align}
\label{eq:Delta_theta_pdf_model}
    P(\Delta\theta;\,\delta) = & \frac{\sin^2\delta}{\pi}\,\left[\frac{\left(\pi-\arccos(\cos 2\Delta\theta\,\cos\delta) \right)\,\cos 2\Delta\theta\,\cos\delta}{\left(1-\cos^2 2\Delta\theta\,\cos^2\delta\right)^{3/2}} + \frac{1}{1-\cos^2 2\Delta\theta \,\cos^2\delta}\right].
\end{align}
The angular parameter $\delta$ sets how much $P(\Delta\theta;\,\delta)$ is peaked at $\Delta\theta=0$. This analytic form is derived under the following general assumptions: (i) normalized integrated Stokes parameters $Q/I$ and $U/I$ are drawn from zero-mean Gaussian distributions and the polarization direction is random; (ii) shape ellipticity variables are drawn from zero-mean Gaussian distributions and the orientation of the ellipse is random; (iii) normalized Stokes parameters and ellipticity variables have a nonzero covariance parameterized by the angle $0\leqslant \delta \leqslant 90^\circ$; the integrated polarization direction aligns perfectly with the semi-minor axis if $\delta=0$, and the two directions are uncorrelated if $\delta=90^\circ$.

We determine the value of $\delta$ for a given sample of $\Delta\theta$ by maximizing the likelihood function. For a given $\delta$ value, the inverse square root of the Fisher information sets the Cram\'{e}r-Rao uncertainty bound on measuring $\alpha$ using any unbiased estimator:
\begin{align}
\label{eq:sigma_alpha_eff}
    \sigma_{\alpha,{\rm eff}} = F^{-1/2} = \left[ \int\rmd\Delta\theta\,P(\Delta\theta;\,\delta)\,\left(\frac{\partial\ln P(\Delta\theta;\,\delta)}{\partial \Delta\theta}\right)^2 \right]^{-1/2}.
\end{align}

Correlation between $\Pi_p$ and the spread of $\Delta\theta$ is physically expected because more inclined (but not nearly edge-on) galaxies have larger polarization fractions~\citep{stil2009integrated} and at the same time show a tighter polarization-shape alignment~\cite{yin2025new}. By Cauchy-Schwarz inequality as in \refEq{Fisher_F}, the Cram\'{e}r-Rao bound can be tightened if we model the joint distribution $P(\Delta\theta,\,\Pi_p)$. With only $\mathcal{O}(10^3)$ selected galaxies, we have insufficient statistics for developing a precise functional form for $P(\Delta\theta,\,\Pi_p)$. Instead, we fit individual inclination bins to the analytic model \refEq{Delta_theta_pdf_model}, so that $P(\Delta\theta,\,\Pi_p)$ is approximated as
\begin{align}
\label{eq:PDelthetap_bins}
    P(\Delta\theta,\,\Pi_p) \approx \sum_I\,w_I\,P(\Delta\theta;\,\delta_I),
\end{align}
where the index $I$ labels inclination bins, $\delta_I$ is the best-fit parameter for each bin, and the weight $w_I$ is the number fraction of each bin satisfying $\sum_I\,w_I = 1$.
Galaxies are sorted into bins according to their apparent inclination in ascending order, such that each bin has approximately the same number of galaxies. In the example of five bins, the first bin has the lowest 20\% of inclinations, the second has the next 20\%, and so on. This ensures comparable galaxy numbers across all bins. The total Fisher information gives the Cram\'{e}r-Rao bound
\begin{align}
    \sigma_{\alpha,{\rm eff}} \approx \left[\sum_{I}\,w_I\,F_I\right]^{-1/2}.
\end{align}
To the extent that \refEq{PDelthetap_bins} well approximates $P(\Delta\theta,\,\Pi_p)$, we expect $\sigma_{\alpha,{\rm eff}}$ calculated this way to improve upon the $\Pi_p$-marginalized PDF model ignoring the inclination dependence. This improvement is feasible because here apparent inclination is used as auxiliary information to weight galaxies.

\begin{figure*}[htbp]
\includegraphics[width=1\textwidth]{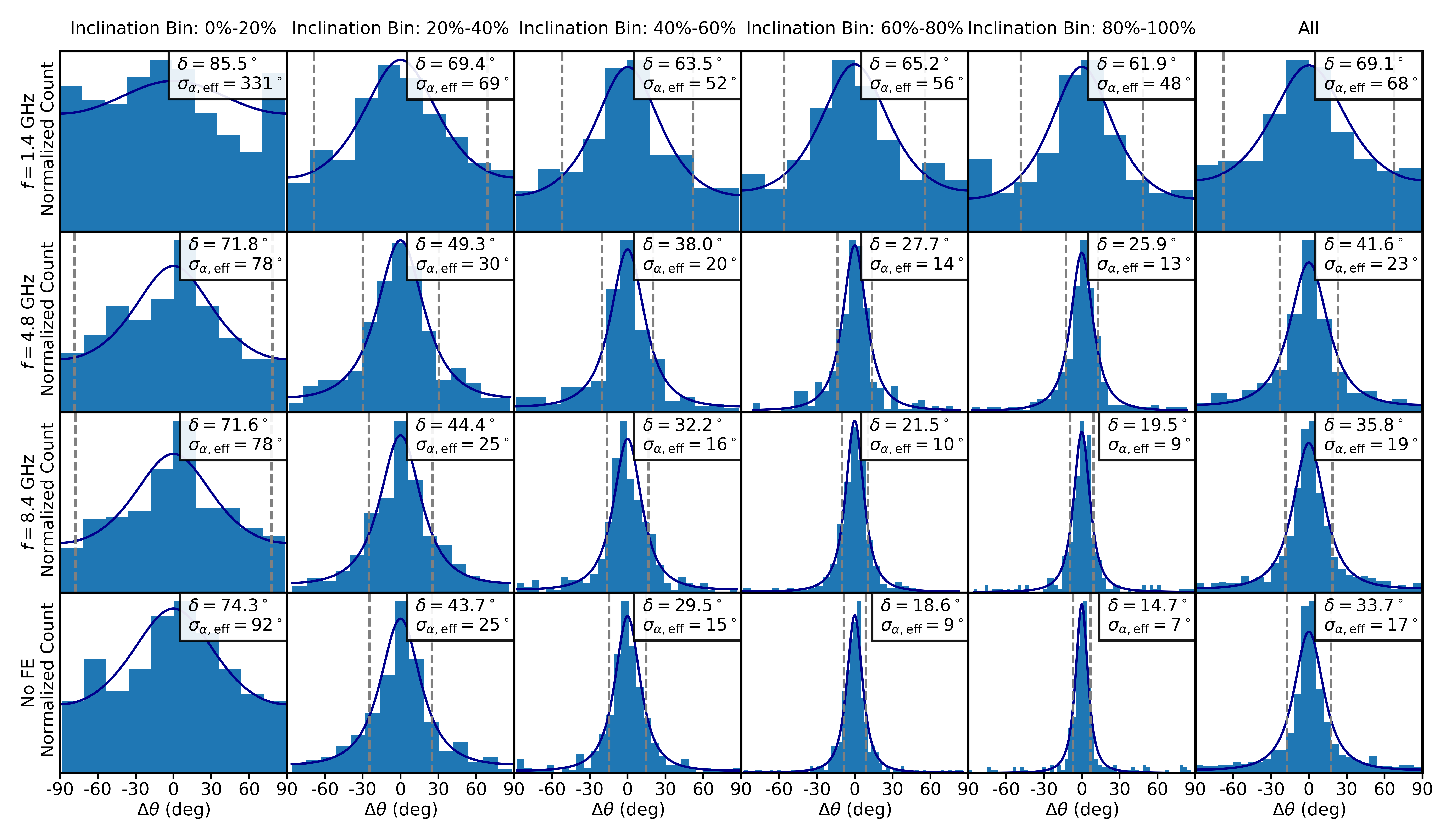}
\caption{Distribution of the polarization-shape misalignment angle $\Delta\theta$ at $z=0$. The rows correspond to 3 different frequencies, $1.4\,$GHz, $4.8\,$GHz and $8.4\,$GHz, compared to the (nearly) frequency-independent case without internal Faraday depolarization (No FE). The columns correspond to statistics in each of the 5 bins into which galaxies are sorted based on the apparent inclination, plus the inclination-inclusive, combined statistics. In each panel, the histogram is derived from our selected $\texttt{IllustrisTNG50}$ galaxies, and the blue curve is the one-parameter analytic fit \refEq{Delta_theta_pdf_model} to the distribution, with the best-fit $\delta$ parameter quoted. The analytic model \refEq{Delta_theta_pdf_model} provides a good fit in individual inclination bins, but when inclination bins are combined the central peak is slightly under-fit. The value of $\sigma_{\alpha,{\rm eff}}$, which gives the effective RMS uncertainty per galaxy in estimating an external polarization rotation angle, is computed from Fisher information (\refEq{sigma_alpha_eff}) and is also quoted.}
\label{fig:delta_fit}
\end{figure*}

When analyzing $N \gg 1$ unrelated galaxies subject to the same amount of external polarization rotation, the overall likelihood is nearly Gaussian, and the uncertainty in $\alpha$ is $\sigma_{\alpha,{\rm eff}}/\sqrt{N}$. Thus, the quantity $\sigma_{\alpha,{\rm eff}}$ computed based on Fisher information is convenient for estimating sensitivity to the polarization rotation angle with many galaxies. We shall refer to it as the {\it effective} RMS misalignment angle.

To illustrate how the effective RMS misalignment angle can be used, consider a sample of $10^4$ galaxies of all inclinations at redshift $z=0$. According to \reffig{delta_fit}, the rotation angle of each galaxy follows the distribution \refEq{Delta_theta_pdf_model} with effective RMS misalignment angle $\sigma_{\alpha,{\rm eff}}=17^\circ$. In the presence of an isotropic cosmic birefringence angle $\alpha$, the mean polarization–shape misalignment angle $\bar{\Delta\theta}$ is, by the central limit theorem, distributed approximately as
\begin{align}
    \bar{\Delta\theta} \sim \mathcal{N}\!\left(\alpha,\, \frac{\sigma^2_{\alpha,{\rm eff}}}{N}\right) 
    = \mathcal{N}\!\left(\alpha,\, (0.17^\circ)^2\right).
\end{align}
Thus, measuring the mean misalignment angle of $10^4$ galaxies at redshift $z=0$ would allow us to constrain an isotropic cosmic birefringence signal with an uncertainty of $\sim0.17\,^\circ$, comparable to current CMB constraints of rotation angle $\alpha = 0.342\,^\circ{}^{+0.094\,^\circ}_{-0.091\,^\circ}$ (68\% CL) \cite{Eskilt:2022cff}, but probing a different redshift range.

\reffig{delta_fit} shows that the analytic model \refEq{Delta_theta_pdf_model} well describes the distribution of $\Delta\theta$ for galaxies at $z=0$, for a range of inclination bins and at different frequencies. The fits perform similarly well at the other redshifts. We tabulate the best-fit $\delta$ values in \reftab{best_fit_eta_vals}, which can facilitate constructing simple semi-analytical models of the misalignment angle statistics.

\begin{table}[htbp]
    \caption{Best-fit $\delta$ parameter for the analytic model \refEq{Delta_theta_pdf_model} for the distribution of polarization-shape misalignment angle $\Delta\theta$, at six redshifts and for three observed frequencies, plus the (nearly) frequency-independent case without internal Faraday effect (No FE). The $\delta$ value is quoted in degrees for each of the five inclination bins as well as for the inclination inclusive sample.}
    \centering
    \begin{tabular}{@{\extracolsep{4pt}}c|c|ccccc|c}
        \hline\hline
        redshift $z$ & frequency [GHz] & bin 1 & bin 2 & bin 3 & bin 4 & bin 5 & all inclinations \\
        \hline
        0.0 & 1.4 & 85.5 & 69.4 & 63.5 & 65.2 & 61.9 & 69.1 \\
        0.0 & 4.8 & 71.8 & 49.3 & 38.0 & 27.7 & 25.9 & 41.6 \\
        0.0 & 8.4 & 71.6 & 44.4 & 32.2 & 21.5 & 19.5 & 35.8 \\
        0.0 & No FE          & 74.3 & 43.7 & 29.5 & 18.6 & 14.7 & 33.7 \\
        \hline
        0.3 & 1.4 & 83.2 & 77.3 & 70.0 & 66.3 & 66.2 & 72.6 \\
        0.3 & 4.8 & 72.5 & 51.6 & 33.2 & 27.9 & 25.5 & 40.8 \\
        0.3 & 8.4 & 71.5 & 51.7 & 30.8 & 22.9 & 17.7 & 37.5 \\
        0.3 & No FE          & 69.9 & 50.2 & 29.6 & 20.5 & 13.2 & 35.5 \\
        \hline
        0.7 & 1.4 & 80.5 & 80.9 & 70.4 & 70.9 & 68.8 & 75.1 \\
        0.7 & 4.8 & 67.7 & 50.8 & 39.7 & 34.7 & 30.6 & 44.1 \\
        0.7 & 8.4 & 65.2 & 49.1 & 34.1 & 28.4 & 23.5 & 39.3 \\
        0.7 & No FE          & 65.8 & 50.1 & 33.6 & 26.3 & 19.8 & 38.5 \\
        \hline
        1.0  & 1.4 & 82.1 & 75.4 & 71.1 & 69.1 & 66.9 & 73.0 \\
        1.0  & 4.8 & 60.3 & 54.1 & 40.9 & 36.6 & 32.9 & 44.6 \\
        1.0  & 8.4 & 59.7 & 53.1 & 36.0 & 32.5 & 28.1 & 41.5 \\
        1.0  & No FE          & 61.2 & 53.1 & 36.4 & 30.6 & 26.0 & 41.3 \\
        \hline
        1.5 & 1.4 & 83.6 & 72.4 & 69.6 & 64.8 & 62.1 & 70.3 \\
        1.5 & 4.8 & 69.0 & 57.9 & 49.7 & 44.8 & 37.9 & 51.6 \\
        1.5 & 8.4 & 68.2 & 51.7 & 48.1 & 41.7 & 35.0 & 49.8 \\
        1.5 & No FE          & 66.9 & 56.0 & 48.7 & 41.5 & 35.6 & 49.5 \\
        \hline
        2.0 & 1.4 & 76.0 & 69.7 & 65.2 & 64.5 & 60.7 & 67.2 \\
        2.0 & 4.8 & 67.3 & 56.0 & 51.2 & 51.0 & 43.1 & 53.6 \\
        2.0 & 8.4 & 67.8 & 54.2 & 50.2 & 50.7 & 42.8 & 53.0 \\
        2.0 & No FE          & 68.6 & 54.9 & 49.9 & 51.4 & 43.1 & 53.6 \\
        \hline\hline
    \end{tabular}
    \label{tab:best_fit_eta_vals}
\end{table}

\begin{figure*}[t]
\centering
    \includegraphics[width=0.495\textwidth]{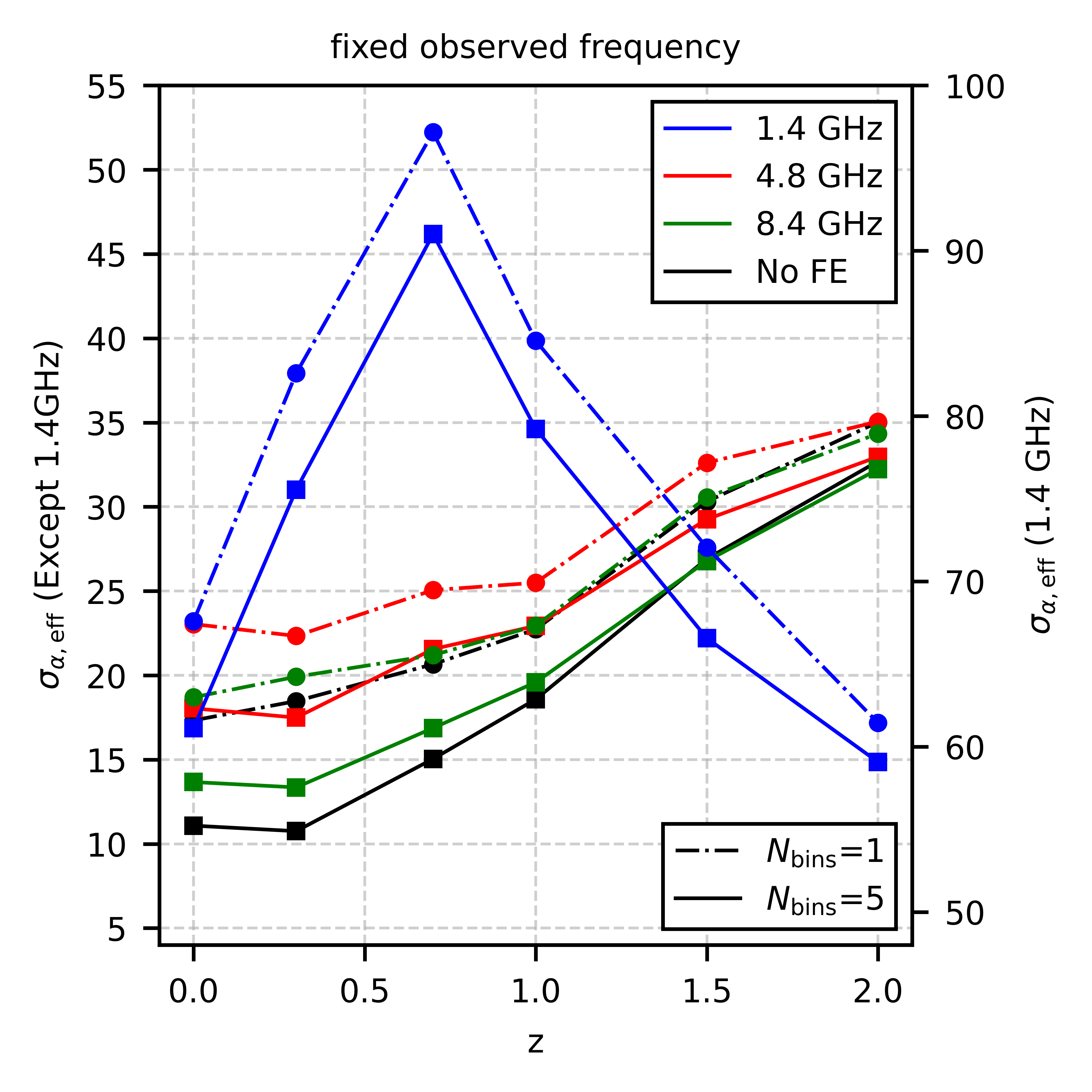}
    \includegraphics[width=0.495\textwidth]{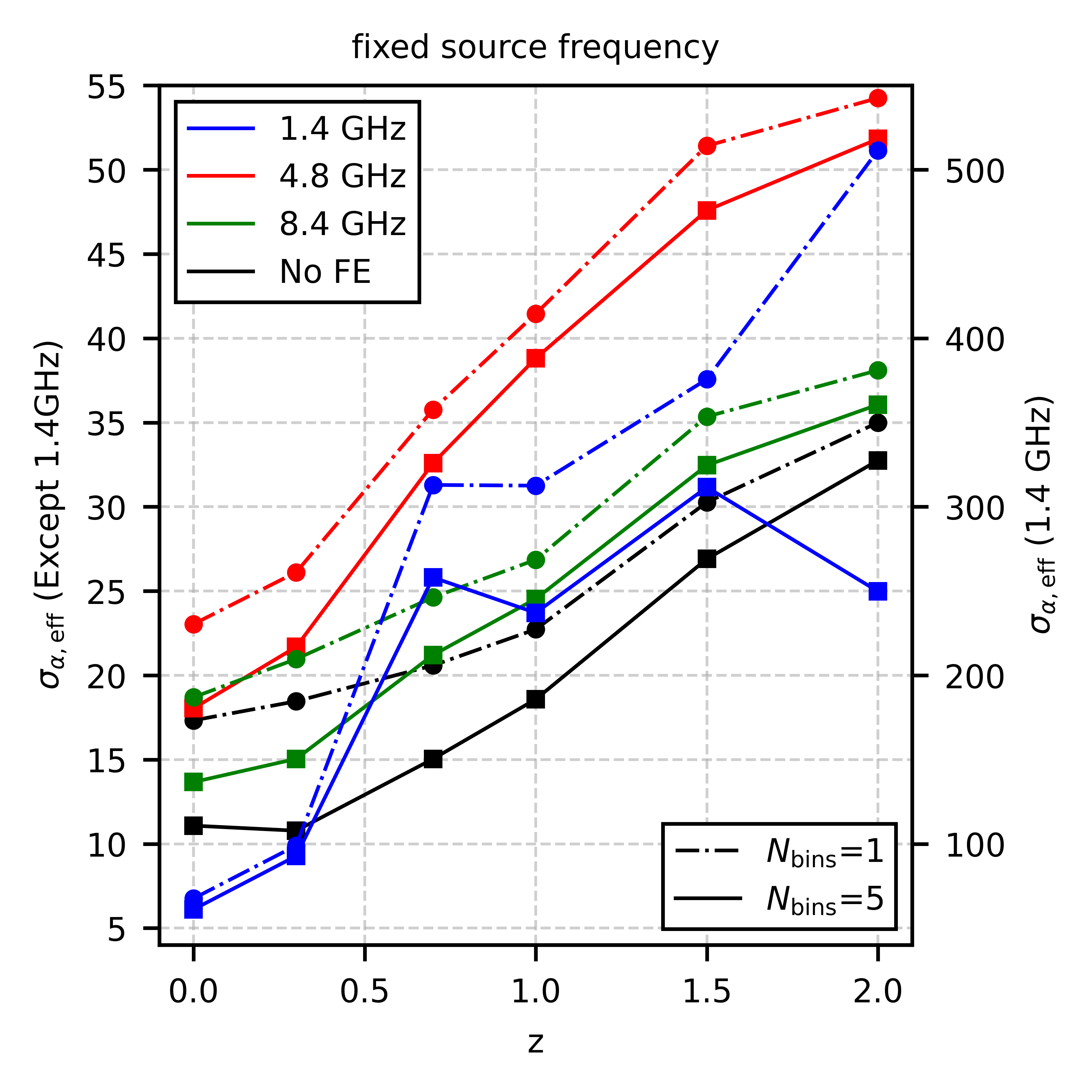}
\caption{Effective uncertainty $\sigma_{\alpha,{\rm eff}}$ per galaxy in estimating the polarization rotation angle $\alpha$ as a function of galaxy redshift. {\bf Left panel:} Results for three fixed {\it observed} frequencies, $1.4\,$GHz, $4.8\,$GHz and $8.4\,$GHz, are plotted separately. They are compared to the calculation without accounting for internal Faraday depolarization (No FE), which is essentially frequency independent. To derive $\sigma_{\alpha,{\rm eff}}$, Fisher information is calculated for both the inclination-inclusive model distribution for the misalignment angle (solid; \refEq{Delta_theta_pdf_model}) and the combination of 5 inclination bins (dash-dotted; \refEq{PDelthetap_bins}). {\bf Right panel:} Same as the left panel, but for fixed frequencies in the {\it source} frame. Polarization-shape alignment degrades with increasing redshift and with decreasing source-frame frequency. When observed at a fixed frequency, reduced internal Faraday depolarization due to the redshifted source-frame frequency partially compensates for the redshift evolution of disk structure and ISM content.}
\label{fig:sig_alpha_eff}
\end{figure*}

Computing Fisher information based on the analytic fit of the misalignment angle distribution, we show in the right panel of \reffig{sig_alpha_eff} the redshift evolution of $\sigma_{\alpha,{\rm eff}}$, at $1.4$, $4.8$ and $8.4\,$GHz in the source frame, respectively. Combining 5 separate inclination bins yields higher Fisher information than modeling the inclination-inclusive statistics. Sorting galaxies into more inclination bins does not yield significantly higher Fisher information, which suggests that the use of 5 bins already adequately exploits the anti-correlation between misalignment and inclination. 

Polarization-shape alignment is seen to degrade rapidly from $z=0$ through $z=2$, already in the (nearly) frequency-independent result neglecting internal Faraday depolarization. This is likely a result of increasingly premature disk structure and irregular ISM magnetic fields toward higher $z$. When Faraday depolarization is accounted for, misalignment angles are too large to be practically useful at $1.4\,$GHz. At $4.8\,$GHz, $\sigma_{\alpha, {\rm eff}}$ rapidly degrades from $18^\circ$ at $z=0$ to $27^\circ$ at $z=0.5$ and to $39^\circ$ at $z=1$. By comparison, Ref.~\cite{yin2025new} report a largely consistent but slightly smaller result at $z=0$, $\sigma_\alpha=5^\circ$--$15^\circ$, by bootstrapping on the small local spiral sample of Ref.~\cite{stil2009integrated} using minimal-variance quadratic estimators. With weaker Faraday depolarization at $8.4\,$GHz, degradation is less rapid with redshift.

For an important point, our results on the misalignment statistics should in fact be interpreted more optimistically given that galaxy are surveyed at a fixed {\it observed} radio frequency. The higher frequency in the source frame is subject to weaker internal Faraday depolarization. As in the left panel of \reffig{sig_alpha_eff}, for observing at $4.8\,$GHz, $\sigma_{\alpha, {\rm eff}}$ only inflates mildly from $18^\circ$ at $z=0$ to $20^\circ$ at $z=0.5$ and to $23^\circ$ at $z=1$. When observing at $8.4\,$GHz, the impact of internal Faraday depolarization is actually insignificant because of cosmic redshifting. Extending the galaxy population to $z=2$, we find $\sigma_{\alpha, {\rm eff}} \simeq 33^\circ$ for $\nu\gtrsim 4.8\,$GHz.

\subsection{Observable galaxy number}

\begin{figure}[t]
\hspace{0\textwidth}
\includegraphics[width=0.6\textwidth]{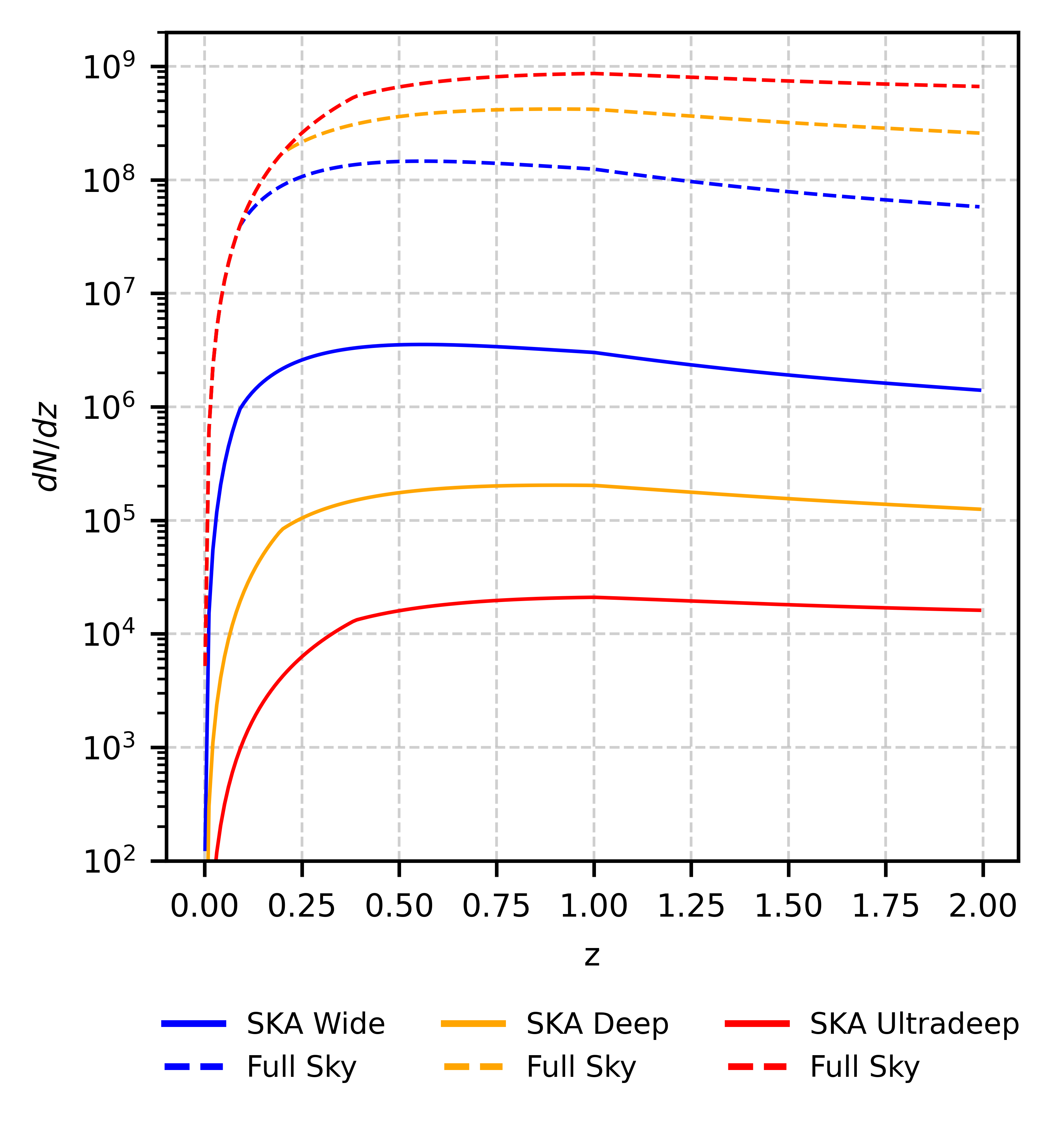}
\centering
\caption{Redshift distribution of observable galaxies in radio continuum surveys for observation at $4.8\,$GHz. The lower cutoff of the specific luminosity is set to be either the SKA benchmark detection limit or $10^{20}\,\mathrm{W\,Hz^{-1}}$, whichever is larger.
\label{fig:galaxy_num_dens}}
\end{figure}

We revisit the number of observable galaxies that can be exploited for such polarization-shape alignment phenomenon. Our calculation is performed using the model RLF of star-forming galaxies from Ref.~\cite{Hansen2024SHARKmodel}. \reffig{galaxy_num_dens} shows the galaxy redshift distribution $\rmd N/\rmd z$. As an example of upcoming radio continuum surveys, we consider three SKA reference surveys, \texttt{Wide}, \texttt{Deep} and \texttt{Ultradeep}, which cover $1000\,{\rm deg}^2$, $20\,{\rm deg}^2$ and $1\,{\rm deg}^2$, and have a $1\sigma$ depth in SKA-MID Band 2 at $\sigma_I=1\,\mu$Jy, $0.2\,\mu$Jy and $0.05\,\mu$Jy, respectively~\citep{Coogan2023SKAStarFormingGalaxies}. Integrating over $z$ from $0.2$ to $1.5$ results in a total of $3.8\,\times\,10^{6}, 2.3\,\times\,10^{5}, 2.3\,\times\,10^{4}$ galaxies whose integrated radio intensities have SNR $>5$ at $4.8\,$GHz. For comparison, integrating over $z$ from $0.8$ to $1.5$ results in a total of $1.9\,\times\,10^{6}, 1.3\,\times\,10^{5}, 1.4\,\times\,10^{4}$ galaxies, respectively. These numbers are similar to what are found in \cite{yin2025new}, which is based on independent literature estimate of the RLF for spiral galaxies.

Limited by integration time, the above reference SKA surveys are supposed to cover a rather small fraction of the sky. Even the \texttt{Wide} survey covers only 2.4\% of the sky. In \reffig{galaxy_num_dens}, we also plot the full-sky galaxy numbers just to show the largest possible sample the Universe can provide in principle. At the depth of the shallowest \texttt{Wide} survey the sample size can increase by 40 times. At the depth of the deepest \texttt{Ultradeep} survey, the sample size can increase by more than four orders of magnitude, enabling reach to cosmic birefringence signals two orders of magnitude smaller. 
If each survey reaches its full potential by covering the entire $41,253\,$deg$^2$ sky, the total available galaxy numbers would increase significantly to $1.6\,\times\,10^{8}, 4.7\,\times\,10^{8}, 9.5\,\times\,10^{8}$ (for $0.2<z<1.5$), and $7.6\,\times\,10^{7}, 2.7\,\times\,10^{8}, 5.8\,\times\,10^{8}$ (for $0.8<z<1.5$), respectively.

\section{Discussion}
\label{sec:discuss}

\subsection{Modeling of radio continuum}

We have neglected thermal free-free emission from H II regions, which also contribute to radio continuum at $1$--$10\,$GHz. Compared to the typical synchrotron spectral slope, thermal free-free emission has a nearly flat flux density $S_\nu$, and is more important toward higher frequencies. In spiral galaxies similar to the MW, thermal free-free emission often has a sub-dominant luminosity than synchrotron for $\nu \lesssim 10\,$GHz~\citep{Klein2018RadioSynchrotron}. In starburst dwarf galaxies, free-free emission can overtake synchrotron already at a few GHz, reflecting high ongoing star formation activity~\citep{Klein2018RadioSynchrotron}. Since thermal free-free emission is unpolarized, including it will not change $Q$ and $U$ and hence the direction of integrated polarization, but will merely reduce the integrated polarization fraction $\Pi_p=\sqrt{Q^2+U^2}/I$. It may be important to include thermal free-free contribution for less mature galaxies that reside in lower-mass dark matter halos and have high specific star-formation rates.

In calculating the integrated emission, we have adopted the simplistic assumption that radio continuum luminosity is locally proportional to star formation rate. While thermal free-free emission is localized to H II regions surrounding young stars, this assumption is only a crude one for synchrotron emission. Since non-thermal electrons, or cosmic rays (CRs), diffuse out from the vicinity of star formation where they are accelerated, synchrotron emission can be better modeled to have a surface brightness profile that is a smoothened version of the far-infrared (FIR) emission profile, which traces instantaneous star formation more closely~\citep{Murphy2006FIRradioCorrelation, Murphy2008FIRRadioCorrCRDiffusion, Vollmer2020RadioKernelSmoothing}. The typical smoothing scale $\sim 1\,$kpc~\citep{Vollmer2020RadioKernelSmoothing} due to CR transport can be larger than the cell resolution of star formation in \texttt{IllustrisTNG50}. The modeling of the spatial distribution of synchrotron emission can be improved with such spatial smoothing. For this reason, the synchrotron emission profile in reality may be less clumpy than what we have modeled here and may trace the symmetric disk geometry better. It may be spatially more extended and hence subject to weaker internal Faraday depolarization. These considerations imply we may have overestimated the variance of the polarization-shape misalignment angle $\Delta\theta$, which should be tested with observations.

\subsection{Implication for cosmological applications}

In order to forecast the angular noise power spectrum of anisotropic polarization rotation angles toward $z<1.5$ spiral galaxies, Ref.~\cite{yin2025new} surmise that the effective RMS misalignment angle is $\sigma_{\alpha, {\rm eff}} = 5^\circ$--$15^\circ$. Inferred from data for a dozen local spirals from Ref.~\cite{stil2009integrated}, this range is subject to large statistical uncertainty and may not apply to galaxies at higher $z$. Moreover, the forecast based on the variance of the quadratic estimator constructed in Ref.~\cite{yin2025new} is mathematically robust only if the spin-0 quantities $X$ and $Y$ have roughly Gaussian distributions. Statistics produced in this work do not support this assumption.

Encouragingly, the Fisher information we calculate here suggests that $\sigma_{\alpha,{\rm eff}}=10^\circ$--$15^\circ$ may be a fairly reliable estimate for galaxies at $z=0$--$0.7$ if observed at 4.8 GHz or higher frequencies. Increase in this number is only modest $\sigma_{\alpha,{\rm eff}}=20^\circ$--$23^\circ$ out to $z=1$. The main conclusions of Ref.~\cite{yin2025new} should therefore largely stand despite the aforementioned limitations.  

The projected performance of SKA-MID continuum surveys will enable efficient buildup of usable disk galaxy sample out to $z=2$. Assuming benchmark SKA-MID survey capabilities, the method will be sample-limited for probing cosmic birefringence sources well beyond $z=2$, but our results suggest that before reaching that redshift, the quality of polarization-shape alignment may have much deteriorated. For cosmic shear measurements, the most accessible redshift range is comparable to current shape surveys such as KiDS~\citep{Giblin2021KiDS1000catalog} and forthcoming ones with Rubin LSST~\citep{Chang2013LSSTweaklensingNeff}, Euclid~\citep{Laureijs2011EuclidDefinitionStudy, Euclid2021OptimizingPhotozSample}, and the Roman Space Telescope~\citep{Wenzl2022RomanCosmologyWeakLensing}. Prevalence of polarization-shape alignment implies that new methods based on polarization-shape alignment have great potential, as in principle there will be $10$--$100$ times more usable galaxies on the sky at the depth of the \texttt{Wide} survey. At the depth of the \texttt{Ultradeep} survey, there will be $10^3$--$10^4$ times more usable galaxies. As benchmark SKA surveys still cover small fractions of the sky, galaxy samples can be substantially enlarged by simply increasing the sky coverage.

\subsection{AGN radio emission}
\label{sec:AGNemission}

We have neglected radio emission powered by an accreting SMBH at the galactic center, which cannot be easily separated from star-formation powered emission when observing galaxies at cosmological distances. Since synchrotron emission powered by AGN is generally polarized, it can change the polarization direction of the integrated emission and hence broaden the distribution of polarization-shape misalignment angle $\Delta\theta$. To see this, imagine a galaxy whose integrated polarization aligns perfectly with the shape minor axis when only radio emission powered by star formation is considered. If additionally the AGN radio emission is, say 10 times fainter, has the same polarization fraction, but has a polarization direction unrelated to the disk, then overall the integrated polarization can become misaligned with the minor axis by up to $0.1\,$radian, or about $6^\circ$. In this way, AGN contamination can degrade polarization-shape alignment.

Since we are unable to reliably predict the AGN radio luminosity and its polarization properties for individual galaxies using \texttt{IllustrisTNG50} simulation data, it is unclear how the $\Delta\theta$ distributions we have derived will degrade once AGN contributions are accounted for, and what fraction of the star-forming galaxy population is subject to significant degradation. Radio-loud AGNs dominate the bright end of the RLF for $\log L_{\nu,{\rm 1.4}}[{\rm W}\,{\rm Hz}^{-1}]>23$. Often residing in quiescent elliptical galaxies, these are orders of magnitude less abundant than star-forming spirals and are not a concern for our purposes.

A significant fraction of the galaxies on the star-forming main sequence like our own MW do not exhibit an AGN. Their central SMBHs are accreting at very low rates and have weak core radio emission that is one order of magnitude or more fainter than disk emission powered by star formation. There is evidence that such core emission has remarkably low polarization fractions $\Pi_p=0.1\%$--$1\%$, such as seen from Sgr A$^*$~\citep{Bower1999SgrAstarVLAspectropolarimetry}, M81~\citep{Brunthaler2001M81AGNpolarization} and M101~\citep{Berkhuijsen2016M101RadioPolarization}, probably due to a combination of frequency-independent and frequency-dependent depolarization effects in the nucleus. The results we derive in this work should be applicable to these ``normal'' galaxies~\citep{Condon1992ARAAreview}.

Other star-forming spiral galaxies do host an AGN, which is often observationally diagnosed through optical spectroscopy. For these so-called Seyfert galaxies, which are one to two orders of magnitude less abundant than non-AGN galaxies, AGN-powered radio emission is certainly non-negligible~\citep{Sebastian2020RadioPolarimetrySeyfert}. For these, the effect of AGN contamination on the polarization-shape alignment is less clear. In these galaxies, the AGN-related nuclear emission can range from dominating the emission from star formation, to being comparable to the star formation contribution, and to being sub-dominant but non-negligible. The overall intrinsic polarization fraction of the AGN contribution is in the range of several percent~\citep{Fanti2004B3VLACSSsamplePolarization} to tens of percent~\citep{Sebastian2020RadioPolarimetrySeyfert}, with an anti-correlation with the linear size of the emission~\citep{Fanti2004B3VLACSSsamplePolarization}. Observations also suggest that fainter AGN emissions tend to be spatially more compact~\citep{Oreinti2014CSSLuminositySize}, and hence may suffer more from depolarization. Given this trend, AGN contamination less than ~10\% of the star-formation powered radio luminosity likely has a low polarization fraction $\Pi_p\lesssim 1\%$ that the $\Delta\theta$ distribution we have derived is not significantly broadened. 

It may be a reasonable estimate that AGN contamination does not significantly impact polarization-shape alignment in at least an order-unity fraction of all star-forming spirals. Ultimately, quantitatively addressing this question requires either data from pilot radio continuum surveys that are deep enough to measure integrated emission from a cosmological population of star-forming galaxies, or cosmological galaxy formation simulations that can reliably predict SMBH-powered radio luminosity and polarization. In practice, it is an interesting question how galaxies having strong AGN-powered emission can be effectively identified and excluded from the sample using optical spectroscopy information.

\section{Conclusion}
\label{sec:concl}

We have carried out a detailed investigation on the phenomenon of polarization-shape alignment in cosmological galaxy formation simulations for the first time. We have derived statistics for the polarization-shape misalignment angle, for thousands of star-forming galaxies at $0 < z < 2$ selected from the \texttt{IllustrisTNG50} simulation. This is done using publicly accessible \texttt{IllustrisTNG50} data, by computing both the optical apparent shape, and the integrated polarized radio continuum emission which is dominated by ISM synchrotron emission energetically sourced by star-forming feedback. To predict the integrated synchrotron emission, we have adopted the simplification that the synchrotron luminosity spatially traces star formation rate perfectly. We have accounted for depolarization by internal Faraday rotation and by randomly oriented small-scale magnetic fields unresolved by simulation cells.

The main findings of this work are summarized as follows:

\begin{enumerate}

\item The derived distributions for the polarization-shape misalignment angle $\Delta\theta$ corroborate the polarization-shape alignment effect in star-forming galaxies over the entire redshift range $0 < z < 2$ we have analyzed. The misalignment angle $\Delta\theta$ is systematically smaller for intermediate to high disk inclinations than for low inclinations, as theoretically expected~\citep{stil2009integrated, yin2025new}.

\item Internal Faraday depolarization strongly degrades polarization-shape alignment. In the source frame, degradation is noticeable already at $8.4\,$GHz, and broadens the distribution of $\Delta\theta$ by more than two-fold at $4.8\,$GHz.

\item The distribution of the polarization-shape misalignment angle $\Delta\theta$ broadens significantly as redshift increases (\reffig{sig_alpha_eff}). This is the case even without accounting for internal Faraday depolarization, reflecting the build-up of geometrically symmetric, thin galactic disks over cosmic time.

\item While the distribution of $\Delta\theta$ is highly non-Gaussian, it is decently fit by the one-parameter analytic form \refEq{Delta_theta_pdf_model}, at all redshifts, in all inclination bins, and at all radio frequencies. We have tabulated the best-fit parameters in \reftab{best_fit_eta_vals} to facilitate analytic modeling. By evaluating Fisher information, we have determined the effective RMS value of $\Delta\theta$, $\sigma_{\alpha,{\rm eff}}$, for estimating external polarization rotation. At $z=0$, we find $\sigma_{\alpha,{\rm eff}}=18^\circ$ at $4.8\,$GHz, similar to local spiral data from Ref.~\cite{stil2009integrated} given sample variance.

\item Since galaxies will be surveyed at fixed observed frequencies, decreased quality in polarization-shape alignment toward larger $z$ is advantageously compensated for by diminished internal Faraday depolarization at the higher source-frame frequencies (\reffig{sig_alpha_eff}). For example, when observed at $4.8\,$GHz, $\sigma_{\alpha,{\rm eff}}$ modestly increases from $18^\circ$ at $z=0$, to $23^\circ$ at $z=1$, and to $33^\circ$ at $z=2$.

\end{enumerate}

A major caveat is that this work has neglected radio emission powered by an accreting SMBH at the galactic center. Such AGN contamination, if bright and sufficiently polarized, can degrade polarization-shape alignment. Strictly speaking, our results only apply to ``normal'' galaxies with negligible radio luminosity from the SMBH, which may still make up an order-unity fraction of all star-forming galaxies. It would be interesting and important to study how distant galaxies with significant AGN emission can be excluded from a radio continuum sample based on auxiliary data such as optical spectroscopy.

Our results motivate efforts to further validate these findings based on the \texttt{IllustrisTNG} simulation data. On the simulation front, it would be useful to see predictions on the polarized radio continuum from zoom-in magneto-hydrodynamic simulations of galaxies with physical treatment of cosmic ray injection and transport (e.g. \citep{Ponnada2024SynchrotronFIRE}). More realistic modeling of the spatial distribution of synchrotron emission may predict improved polarization-shape alignment. While unpolarized and often sub-dominant at several GHz, thermal free-free continuum from H II regions should also be included, perhaps crucially for lower-mass galaxies with high specific star formation rates. Simulations that allow reliable prediction of polarized AGN-powered emission will be highly useful.

On the observational side, radio polarimetry samples of low-$z$ galaxies that reach the ``radio-quiet'' regime $21 < \log L_{\nu, 1.4}[{\rm W}\,{\rm Hz}^{-1}] < 23$ and have adequate sizes $\mathcal{O}(10^2$--$10^3)$ need to be compiled to empirically confirm the findings of this work. This will account for AGN contamination too. For example, the MIGHTEE continuum survey may be deep and large enough to test the alignment effect at $1.2$--$1.3\,$GHz~\citep{Hale2025MIGHTEEsurveyDR1}, even though we predict that internal Faraday rotation significantly weakens polarization-shape alignment at those frequencies. The VLA-COSMOS survey at $3\,$GHz~\citep{Smolcic2017VLACOSMOS3GHzLargeProject} is another existing dataset that targets a more interesting frequency.

This work motivates more detailed study of new robust statistical methods to measure cosmic birefringence and/or cosmic shear using the combination of radio polarimetry and galaxy imaging. For cosmic birefringence in particular, Ref.~\cite{yin2025new} showed that a probe based on polarization-shape alignment can be more sensitive to spatially inhomogeneous sources of polarization rotation at $z<1$ than even Stage-IV CMB experiments~\citep{CMB-S4:2016ple}. Leveraging radio continuum surveys for this science will have great value in the face of an uncertain timeline for Stage-IV CMB programs. 

\acknowledgments
The authors would like to thank Neal Dalal, Kangning Diao, Matthew Johnson, Christopher McKee, Kendrick Smith and Jeroen Stil for useful discussions. R.Z. acknowledges the Berkeley Global Access (BGA) program, which enables the visitor scholarship during which this research was initiated. L.D. acknowledges research grant support from the Alfred P. Sloan Foundation (Award Number FG-2021-16495), from the Frank and Karen Dabby STEM Fund in the Society of Hellman Fellows, and from the Office of Science, Office of High Energy Physics of the U.S. Department of Energy (Award Number DE-SC-0025293). Research at Perimeter Institute is supported in part by the Government of Canada through the Department of Innovation, Science and Economic Development Canada and by the Province of Ontario through the Ministry of Economic Development, Job Creation and Trade. 

\appendix

\section{Galaxy shape measurement}
\label{app:galaxy_shape}

Assuming that each galaxy appears as an ellipse on the sky, we evaluate the ellipticity position angle $\theta_e$ according to 
\begin{align}
\label{eq:theta_e_def}
    \theta_{e} = -\frac{1}{2} \arctan\left(\frac{2\,M_{xy}}{M_{xx} - M_{yy}}\right) + \frac{\pi}{2},
\end{align}
With the plane of sky parameterized by the Cartesian coordinates $x$ and $y$, the photometric second moments $M_{xx}$, $M_{xy}$ and $M_{yy}$ are evaluated from
\begin{align}
M_{xx} &= \frac{\sum l\, x^{2}}{\sum l} - x_c^2, \\
M_{xy} &= \frac{\sum l\, x\,y}{\sum l} - x_c\,y_c, \\
M_{yy} &= \frac{\sum l\, y^{2}}{\sum l} - y_c^2,
\end{align}
where $x_c$ and $y_c$ are coordinates of the photometric center
\begin{align}
    x_c = \frac{\sum l\, x}{\sum l}, \quad y_c = \frac{\sum l\, y}{\sum l}.
\end{align}
Here we sum over simulation cells, and $l$ is the stellar luminosity in the chosen photometric band in each cell, which is extracted from \texttt{IllustrisTNG50} simulation data.

The semi-major axis $a$ and the semi-minor axis $b$ of the apparent galaxy shape are also determined from the second photometric moments
\begin{align}
    M =
    \begin{pmatrix}
        M_{xx} & M_{xy} \\
        M_{xy} & M_{yy}
    \end{pmatrix}.
\end{align}
Then $a$ and $b$ are the square roots of the two eigenvalues of matrix $M$, $\lambda_1$ and $\lambda_2$, with $\lambda_1 > \lambda_2$,
\begin{align}
    a = \sqrt{\lambda_1}, \qquad b = \sqrt{\lambda_2}.
\end{align}
The galaxy inclination $i$ is then estimated as
\begin{align}
    i = \arccos(b/a).
\end{align}
It should be noted that this thin-disk approximation neglects the presence of a galactic bulge and does not account for the intrinsic finite thickness of disk galaxies. A more general expression introduces a minimum intrinsic axial ratio $q_0$ \cite{Holmberg1958APP},
\begin{align}
    \cos^2 i = \frac{(b/a)^2 - q_0^2}{1 - q_0^2},
\end{align}
where $q_0$ depends on morphological type. Our analysis does not adopt this correction. But when $q_0 = 0$, this expression reduces to the thin-disk approximation used above.

\section{Polarized synchrotron emissivity}
\label{app:sync_emis}

While many analytic results presented in this Appendix can be found in textbooks on electrodynamics or radiative processes (e.g. \cite{jackson2021classical, rybicki2024radiative}), we collect them here to ensure the discussion of this paper is self-contained. 

Assuming there is a power-law distribution of relativistic electrons with respect to the Lorentz factor $\gamma$
\begin{align}
    \rmd n/\rmd\gamma = n_0\,\gamma^{-p}, \quad \gamma_{\rm min} < \gamma < \gamma_{\rm max}.
\end{align}
Here $n$ is the number density of non-thermal electrons, $n_0$ is a normalizing factor, $p = 4.4$ is a typical spectral index for the ISM~\cite{Marinacci2018IllustrisTNGradiohaloBfields}, $\gamma_{\rm min} = 300$ is the lower energy cut-off for non-thermal electrons. The upper energy limit is chosen to be $\gamma_{\rm max} = 10^5$ instead of infinity to save computation~\cite{westfold1959polarization, Marinacci2018IllustrisTNGradiohaloBfields}.

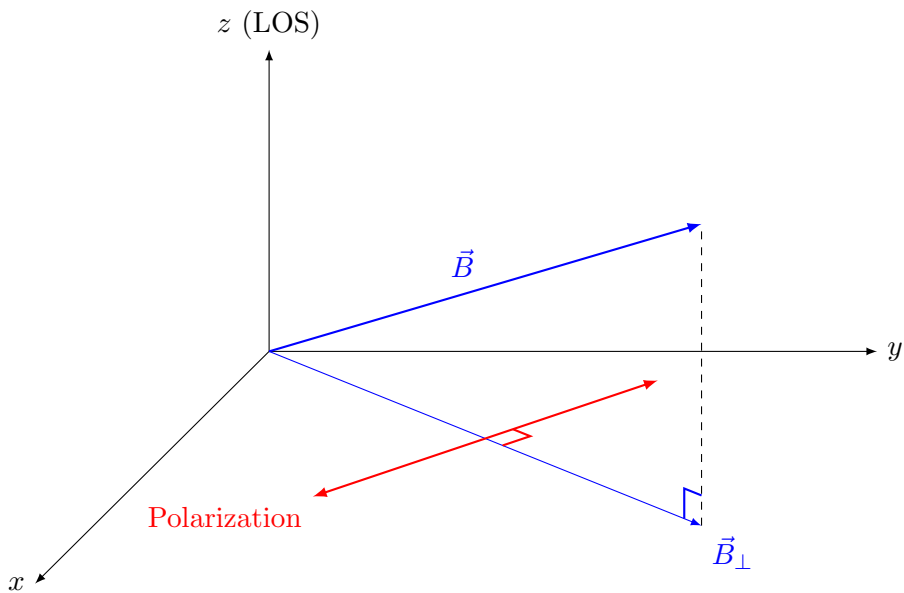
\begin{figure}[htbp]
\centering
\begin{tikzpicture}[scale=2, >=latex]

  \draw[->] (0,0,0) -- (4,0,0) node[right] {$y$};
  \draw[->] (0,0,0) -- (0,2,0) node[above] {$z$ (LOS)};
  \draw[->] (0,0,0) -- (0,0,4) node[left] {$x$};

  \draw[->, thick, blue] (0,0,0) -- (4,2,3) node[midway, above left] {$\vec{B}$};

  \draw[->, blue] (0,0,0) -- (4,0,3) node[below right] {$\vec{B}_\perp$};
  \draw[dashed, black] (4,0,3) -- (4,2,3);

  \draw[-, thick, blue] (4-0.16,0,3-0.12) -- (4-0.16,0+0.2,3-0.12) -- (4,0+0.2,3);

  \draw[<->, thick, red] (1.25,0,2.5) node[below left] {Polarization} -- (2.75,0,0.5);
  
  \draw[-, thick, red] (2+0.06*2,0,1.5-0.08*2) -- (2+0.14*2,0,1.5-0.02*2) -- (2+0.08*2,0,1.5+0.06*2);

\end{tikzpicture}
\caption{Geometry of synchrotron polarization. The line of sight is along the $z$ axis. The projected magnetic field $\vec{B}_\perp$ lies in the sky plane ($x$–$y$ plane), and the observed polarization is perpendicular to $\vec{B}_\perp$ in this plane. Here the subscript ``$\perp$'' denotes the component perpendicular to the line of sight.}
\label{fig:polarization_geometry}
\end{figure}

For a given magnetic field, the observed synchrotron polarization is oriented perpendicular to the projection of the magnetic field onto this plane~\cite{westfold1959polarization} (see \reffig{polarization_geometry} for an example). In a coordinate system aligned with the transverse component of the magnetic field, the Stokes  parameters are
\begin{align}
    I & \propto G(f / f_{\rm max}) - G(f / f_{\rm min}), \\
    Q & \propto G_p(f / f_{\rm max}) - G_p(f / f_{\rm min}),
\end{align}
while $U$ and $V$ vanish. The two functions $G(x)$ and $G_{p}(x)$ are given by
\begin{align}
    G_{p}(x) & = \int_{x}^{\infty} \xi^{(p - 1) / 2}\,K_{2/3}(\xi)\, \rmd\xi, \\
    G(x) & = \frac{p + 7/3}{p + 1}\,G_p(x) - \frac{2\,x^{\frac{p - 1}{2}}}{p + 1}\,\left[ F(x) - F_p(x) \right],
\end{align}
where
\begin{align}
    F(x) = x\,\int_{x}^{\infty} K_{5/3}(\xi)\,\rmd\xi, \quad F_{p}(x) = x\,K_{2/3}(x). 
\end{align}
Here $K_{\nu}(x)$ is the modified Bessel function of the second kind.

The two frequency constants, $f_{\rm min}$ and $f_{\rm max}$, are defined as
\begin{align}
    f_{\rm min} = \frac{3}{2}\,\gamma_{\rm min}^2\,f_{B} \sin{\theta},  \\
    f_{\rm max} = \frac{3}{2} \,\gamma_{\rm max}^2\,f_{B} \sin{\theta}, 
\end{align}
where $f_{B} = {e\,B}/{2 \pi \,m_e}$ is the gyrating frequency. The pitch angle $\theta$ is the angle between the line of sight and the magnetic field. Neglecting synchrotron self-absorption and free-free absorption at $1$--$10\,$GHz, this calculation gives the polarized emission from any single volume with a uniform magnetic field.

Coordinate transformations need to be applied to the Stokes parameters from different cells before their contributions are summed up. Let $\theta_B$ be the angle between the reference direction on the sky in radio observation and the transverse component of the magnetic field. The observed Stokes parameters are given by (e.g. \cite{Ponnada2024SynchrotronFIRE})
\begin{align}
    I' = I, \quad Q' = Q\,\cos{2 \theta_{B}}, \quad U' = - Q\,\sin{2 \theta_{B}}.
\end{align}
Summing over all simulation cells, the integrated emission has a polarization position angle $\theta_{p}$ and a polarization fraction $\Pi_{p}$, which are given by
\begin{align}
    \theta_{p} & = -(1/2)\,\arctan\left(U_{\rm tot}/Q_{\rm tot}\right) + \pi/2, \\
    \Pi_{p} & = \sqrt{Q^2_{\rm tot} + U^2_{\rm tot}}/I_{\rm tot}.
\end{align}
Here $\theta_p$ is measured relative to the reference position on the sky.

\section{Electron number density}
\label{app:ne}

The electron number density $n_e$ must be calculated separately for star-forming and non-star-forming cells in \texttt{IllustrisTNG50}~\footnote{Non-star-forming cells are defined as those with $\mathrm{SFR} = 0\ M_{\odot}\,{\rm yr}^{-1}$, i.e., below the SFR resolution limit $10^{-5}\ M_{\odot}\,{\rm yr}^{-1}$ in the simulation.}.
For non-star-forming cells, the ratio between electron number density and hydrogen number density is given in the simulation data as the \texttt{Electron Abundance} entry, $e_{ab}$, and we obtain $n_e$ using the hydrogen number density $n_{\rm H}$~\citep{walker2024dispersion}
\begin{align}
    n_e = e_{ab}\,n_{\rm H}.
\end{align}
Here $n_{\rm H}$ is simply derived from the gas mass density $\rho$
\begin{align}
    n_{\rm H} = X_{\rm H}\, \rho / m_p,
\end{align}
where $m_p$ is the proton mass and we set the hydrogen fraction $X_{\rm H} = 0.76$.

For star-forming cells, we assume that only a fraction of the gas are hot and ionized. This mass fraction, $x_{\rm warm}$, is derived from an average gas temperature $T$ that describes each cell.
\begin{align}
    x_{\rm warm} = \frac{T - T_{\rm cold}}{T_{\rm hot} - T_{\rm cold}}
\end{align}
This temperature $T$ is not directly provided in the simulation data, but we can use the provided \texttt{Internal Energy} entry to derive it. Let $u$ be the amount of thermal energy per unit of unit gas mass in a given cell, which is provided as the \texttt{Internal Energy} entry. We then calculate
\begin{align}
    T = \frac{\mu\,(\gamma - 1)}{k_B}\,u,
\end{align}
where $\gamma = 5/3$ is the adiabatic index for monatomic gas, and $\mu$ is the mean molecular weight
\begin{align}
    \mu = \frac{4\,m_p}{1 + (3 + 4\,e_{ab}) \, X_{\rm H}}.
\end{align}
Hot gas and cold gas are assumed to have fixed temperatures $T_{\rm cold}=10^3\,$K and $T_{\rm hot}=10^7$\,K, respectively~\citep{walker2024dispersion}.
After obtaining the mass fraction of the warm gas, $x_{\rm warm}$, the electron number density for star-forming cells is given by
\begin{align}
    n_e = x_{\rm warm} \, (X_{\rm H} + 2\,Y_{\rm He} / 4 )\, \rho / m_p
\end{align}
In this way, we derive $n_e$ for all cells, which is used in calculating the Rotation Measure (RM) in order to characterize internal Faraday depolarization effects.

\section{Sub-grid model for unresolved magnetic field}
\label{app:subgridBfields}

To more realistically model integrated synchrotron polarization, we adopt a model in which the ISM magnetic field is decomposed into a regular field component $\mathbf{B}_{\rm reg}$ and a random field component $\mathbf{B}_{\rm rand}$. The regular field component is spatially ordered and is defined to be resolved at the level of simulation cells. Within each cell, this component is assumed to be uniform. On the other hand, the random field component is unresolved in the simulation and must be modeled analytically as sub-grid physics. We assume that this component has a coherent length scale much smaller than the cell size and has an isotropic distribution of field direction. Since synchrotron polarization direction is determined by the direction of the local magnetic field, having a spatially unresolved random field component decreases the spatially integrated polarization.

We assume that on average the random field component $\mathbf{B}_{\rm rand}$ has a constant magnitude $r$ relative to the regular field $\mathbf{B}_{\rm reg}$. The integrated polarized synchrotron emissivity tensor $\eta_{ij}$ (which encodes the Stokes parameters) is proportional to the second moment of the transverse magnetic field (the line-of-sight magnetic field does not contribute to the instantaneous acceleration of the electron, as synchrotron radiation is ultra-relativistically beamed along the line of sight)
\begin{align}
    \eta_{ij} \propto \langle B_{i,\perp} B_{j,\perp} \rangle,
\end{align}
which involves the total transverse magnetic field
\begin{align}
    B_{\perp} = B_{{\rm reg},\perp} + B_{{\rm rand},\perp}.
\end{align}
The observed polarization fraction is
\begin{align}
    \Pi_p(r) = \frac{\eta_{xx} - \eta_{yy}}{\eta_{xx} + \eta_{yy}} = \frac{B_{{\rm reg},x}^2 - B_{{\rm reg},y}^2}{(B_{{\rm reg},x}^2 + B_{{\rm reg},y}^2) + (2/3)\,r^2\,B_{{\rm reg}}^2}
\end{align}
For isotropic distributions at the ensemble level, $\langle B_{{\rm reg},x}^2\rangle = \langle B_{{\rm reg},y}^2 \rangle = \langle B_{{\rm reg},z}^2\rangle = (1/3)\,B_{\rm reg}^2$ and 
$\langle B_{{\rm rand},x}^2 \rangle = \langle B_{{\rm rand},y}^2 \rangle = \langle B_{{\rm rand},z}^2 \rangle = (1/3)\,\langle B_{\rm rand}^2 \rangle = (1/3)\,r^2\,B_{\rm reg}^2$, we have
\begin{align}
    \left\langle \Pi_p(r)/\Pi_p(0) \right\rangle \simeq 1/(1 + r^2).
\end{align}
Therefore, in our simple treatment, the integrated polarization is reduced by a constant factor $1/(1+r^2)$ due to sub-grid random magnetic fields.

The value of the ratio $r$ can be inferred from the fact that the random to total magnetic field ratio has the typical value $\langle B_{\rm rand}/B_{\rm tot}\rangle =0.6$--$0.7$ in the Milky Way~\citep{Beck2004role}. In our model, 
\begin{align}
    B_{\rm tot} = B_{\rm rand}\,\sqrt{1 + r^2 - 2r \cos\theta},
\end{align}
where the angle $\theta$ is isotropically distributed. Therefore, we have
\begin{align}
    \left\langle \frac{B_{\rm rand}}{B_{\rm tot}} \right\rangle = \frac{1}{2} \int_{-1}^{1} \frac{\rmd \mu}{\sqrt{1 + r^2 + 2\,r\,\mu}} = \frac{1}{2\,r} \ln\frac{|1 + r|}{|1 - r|}.
\end{align}
We numerically solve this the equation to find that $\langle B_{\rm rand}/B_{\rm tot} \rangle = 0.6$ ($0.7$) corresponds to $r = 1.435$ ($1.354$) and $\langle \Pi_p(r)/\Pi_p(0) \rangle = 0.327$ ($0.353$). Including the unresolved random magnetic fields suppress the polarization fraction by roughly a factor of three.


\bibliographystyle{JHEP}
\bibliography{main}

\end{document}